\documentclass[referee]{raa}           

\usepackage{graphicx,times}
\usepackage{natbib}
\usepackage{amssymb,amsmath}
\bibpunct{(}{)}{;}{a}{}{,}

\usepackage[pagebackref=true]{hyperref}

\begin{document}

   \title{On the spatial distribution of luminous blue variables in the M33 galaxy}

 \volnopage{ {\bf 20XX} Vol.\ {\bf X} No. {\bf XX}, 000--000}
   \setcounter{page}{1}

   \author{A.~Kostenkov
      \inst{1}
   \and S.~Fabrika
      \inst{1}
   \and A.~Kaldybekova
      \inst{2}
   \and S.~Fedorchenko
      \inst{3}
   \and Y.~Solovyeva
      \inst{1}
   \and E.~Dedov
      \inst{1}
   \and A.~Sarkisyan
      \inst{1}
   \and A.~Vinokurov
      \inst{1}
   \and O.~Sholukhova
      \inst{1}
   }

   \institute{ Special Astrophysical Observatory, Nizhnij Arkhyz 369167, Russia; {\it kostenkov@sao.ru}\\
        \and
             Space Research Institute, Russian Academy of Sciences, 117997, Moscow, Russia\\
        \and
             Department of Informatics, University of Z\"{u}rich, Binzm\"{u}hlestrasse 14, Z\"{u}rich CH-8050, Switzerland\\
\vs \no
   {\small Received 20XX Month Day; accepted 20XX Month Day}
}

\abstract{In the current paper, we present a study of the spatial distribution of luminous blue variables (LBVs) and various LBV candidates (cLBVs) with respect to OB associations in the M33 galaxy. The identification of blue star groups was based on the LGGS data and was carried out by two clustering algorithms with initial parameters determined during simulations of random stellar fields. We have found that the distribution of distances to the nearest OB association obtained for the LBV/cLBV sample is close to that for massive stars with $M_{\rm init}>20\,M_\odot$ and Wolf-Rayet stars. This result is in good agreement with the standard assumption that luminous blue variables represent an intermediate stage in the evolution of the most massive stars. However, some objects from the LBV/cLBV sample, particularly Fe\,II-emission stars, demonstrated severe isolation compared to other massive stars, which, together with certain features of their spectra, implicitly indicates that the nature of these objects and other LBVs/cLBVs may differ radically.
}

   \authorrunning{A.~Kostenkov et al.}            
   \titlerunning{On the spatial distribution of luminous blue variables in the M33 galaxy}  
   \maketitle

%
\section{Introduction}           

Luminous blue variables (LBVs) are a rare type of luminous stars with spectral and irregular photometric variability \citep{Humphreys1994, Weis2020}. According to the standard view on the evolution of single stars, LBVs represent the transitional stage between massive O-stars ($\gtrsim$ 25 M$_\odot$) and Wolf-Rayet stars (Conti scenario, \citet{Conti1975, Conti1984}). It corresponds to the transition from hydrogen core burning to helium core burning, when stellar wind and eruptions of LBVs play a defining role in the removal of the hydrogen envelope and the formation of Wolf-Rayet stars \citep{Groh2014}. LBVs are often considered to be an immediate precursor of core-collapse supernovae \citep{Groh2013,Trundle2008, Andrews2021}.

A relatively new approach to explaining the LBV phenomenon is to consider these stars as a result of close binary evolution. \citet{Smith2015} found that 18 LBVs and 11 LBV candidates in the Magellanic Clouds are far more dispersed than massive O-type stars. According to \citet{Smith2015}, the projected distances between LBVs and their first or second nearest O-star are much greater than would be expected for a single massive star. The authors suggested that the isolation of LBVs can be explained by the evolution of close binary stars, and LBVs represent stars "rejuvenated" by accretion. However, \citet{Humphreys2016} pointed out that \citet{Smith2015} used a mixed star sample that includes standard LBVs (M$_\text{init} \gtrsim$ 50 M$_\odot$), low luminosity LBVs (M$_\text{init} \approx$ 25--40 M$_\odot$), LBV candidates (cLBVs), and also eruptive $\eta$~Car-like LBVs, which have different evolutionary status. Then \citet{Smith2016} once again compared distance distributions between LBVs and O-stars, taking into account the notes of \citet{Humphreys2016}, and the previously discovered isolation was confirmed. The authors showed that the most luminous LBVs have an environment similar to that of late O-type stars, whose average age is about twice as large as the expected age of these LBVs in the single massive star evolution scenario. In addition, \citet{Smith2016} also found that confirmed low luminosity LBVs have a spatial distribution close to that of red supergiants with initial masses of M$_\text{init}\sim10-15\,$M$_{\odot}$, which is about half the mass of low luminosity LBVs.

\citet{Aadland2018} revised the photometric criteria for selecting the most massive unevolved stars in the Large Magellanic Cloud (LMC), M31 and M33 galaxies. The authors concluded that the spatial distribution of LBVs/cLBVs (total of 78 objects in three galaxies) is close to that of bright blue stars (BBSs), while older Wolf-Rayet stars and red supergiants (RSGs) are more dispersed. However, \cite{Smith2019} pointed out the inconsistency of the BBS sample in the study by \cite{Aadland2018}, which includes a large number of B-supergiants with masses of $\sim20\,\text{M}_\odot$. \cite{Kraus2019} also noted that the sample used in \cite{Aadland2018} contains B[e]-supergiants classified as LBV candidates, which distorted the result. \cite{Smith2019} concluded that the stellar population selected by \cite{Aadland2018} corresponds to early B-supergiants aged about 10 million years, rather than the youngest and most massive O-stars aged about 3-4 million years. The author notes the similarity between the spatial distributions of LBVs and early B-supergiants, which are less massive and older than LBVs, and explains this phenomenon by the evolution of LBVs in a binary system \citep{Aghakhanloo2017}. 

The calculation of models of young stellar clusters spatial distribution \citep{Aghakhanloo2017} showed that the standard model of single star evolution is not consistent with the stellar environment of LBVs in most cases. To explain the phenomenon of LBVs isolation, authors examined two possible models of their evolution in close binaries. In the first model, LBVs are the result of the merging of two massive stars \citep{Justham2014}, and their isolation is explained by the increased maximum lifetime of the star and consequently a longer time to move away from other massive stars. An alternative model considers the mass transfer to a less massive star in a binary system and the subsequent explosion of a more massive star as a supernova with the ejection of a material-rich companion from the system (binary supernova scenario, BSS) \citep{Blaauw1961}. 

Thus, it is still debated whether the LBV is a evolved single massive star or product of evolution in a binary system. \cite{Humphreys2016, Aadland2018} showed that the conclusions about LBV isolation made by \cite{Smith2015, Smith2016, Smith2019} are tightly dependent on the sample of OB stars. The reliability of the results is also affected by the small number of known LBVs and LBV candidates.

An investigation of the projected distribution of the stellar population in the M33 galaxy by \cite{Bastian2007} showed that OB stars exhibit a hierarchical structure of star formation regions: some stars, probably the youngest ones, form dense groups at the smallest spatial scales that may be included in extended regions with lower concentration of aged blue stars. In fact, several clustering algorithms that identify dense stellar groups have been used successfully by many authors to search for OB associations \citep{Battinelli1991, Ivanov1996, Gouliermis2000, Borissova2004, Chemel2022}. Such papers have led us to the idea that studying the distribution of projected distances to the center of the nearest group of blue stars would allow us to draw more definite conclusions about the evolutionary status of LBVs, compared to studying the distribution of distances to the first or second nearest massive star, due to a less influence of statistical errors and effects related to blue star sample heterogeneity. If LBVs are the evolutionary stage of a single massive star, they are expected to be spatially close to the projected centers of young stellar clusters consisting predominantly of massive blue stars, and their projected distribution with respect to OB associations should be similar to that of O-type or WR stars. The current work is focused on studying spatial distribution of LBVs and potentially evolutionary related stars using the M33 galaxy as a case study. The choice of the galaxy was due to a relatively large number of known LBV and LBV candidates, its spatial location, and availability of high-quality photometric data.

The text of the article is organized as follows: section~\ref{methods} presents the photometric data samples used in the study, various methods of searching stellar associations, and results of studying model data using the selected methods; in section~\ref{results}, we show the distribution of the distances of different types of stars, including LBVs/cLBVs, with respect to nearest identified groups of OB stars, the obtained results are compared with studies of other authors; section~\ref{discussion} focuses on the analyzing obtained spatial distributions in the context of possible scenarios of LBVs evolution.

\section{Methods} \label{methods}

In the first stage of the work, we analyzed the photometric criteria for sampling of the massive stars presented in the studies of other authors. Next, we selected two groups of stars in accordance with chosen color indices and luminosities. The first sample included the most luminous hot stars, which are presumably the evolutionary precursors of LBVs in the single-evolution scenario. The second group consists of blue stars with lower luminosities and, accordingly, lower initial masses. This separation allowed us to make a comparative analysis of the spatial distribution of stars of different ages.

In the second part of the study, we explored different clustering methods that can be used to search for OB associations. The optimal parameters of the selected algorithms were determined using modeling of the location of stars in the picture plane, which was performed based on the information about the stellar environment of known LBV/cLBV in the galaxy M33. At the final stage, we plotted distributions of distances from massive stars to centers of the nearest identified cluster and performed a comparative analysis of projected positions of stars of different types with respect to OB associations.

\subsection{Photometric data} \label{methods_selection}

\begin{figure*}[t!]
       \centering
       \includegraphics[width=0.65\linewidth]{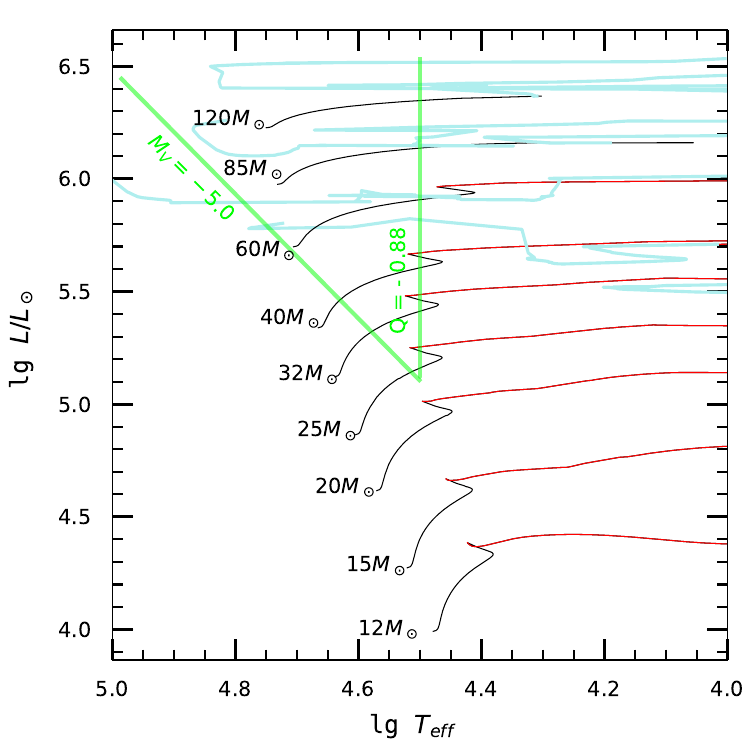}
       \caption{Hertzsprung-Russell diagram with evolutionary tracks for Z=0.006 \citep{Eggenberger2021}. Black lines of the evolutionary tracks represent the main sequence stage, colored --- later stages of the stellar evolution followed after the end of hydrogen burning in the core (red) and in the shell (cyan). Green lines show the region of the diagram with the most massive unevolved stars. The color to the effective temperature $Q-T_{\text{eff}}$ transformation  and bolometric corrections are calculated according to the relations presented in \cite{Massey1995b} and \cite{Massey2005} correspondingly.}
       \label{fig1}
\end{figure*}

Similar to the investigation of the LBV environment presented in \cite{Aadland2018}, our study uses LGGS UBVRI photometry of stars in the M33 galaxy \citep{Massey2006, Massey2016}. For obtaining the sample of most massive stars, \cite{Aadland2018} applied a magnitude restriction criterion of $M_V < -5.0$  that includes interstellar reddening of $E(B-V)=0.13$ typical for OB stars in the M33 galaxy \citep{Massey1995a, Massey2007}. In addition, \cite{Aadland2018} used the reddening-free index ${Q=(U-B)-q_r\times(B-V)}$ with standard value slope of the reddening curve ${q_r=E(U-B)/E(B-V)\approx0.72}$ for the selection of the hottest stars \citep{Massey1995a}.
The sample was restricted by $Q<-0.88$, which corresponds to the effective temperature of $T_{\text{eff}}>35000\,$K for supergiants \citep{Massey1995b}. Moreover, authors adopted additional criteria of $Q<-1.20$, ${(U-B)<-0.5}$ and ${(B-V)<0.2}$ to exclude stars with unrealistic colors and large photometric errors. Fig.\,\ref{fig1} shows the approximate borders of the Hertzsprung–Russell diagram matching the aforesaid photometric criteria along with evolutionary tracks. This figure is identical to figure~1 from the study by \cite{Aadland2018}, however, we employ evolutionary tracks with subsolar metallicity Z=0.006 \citep{Eggenberger2021} according with the oxygen abundance in the H\,II areas of the M33 galaxy \citep{Kang2012}.

\cite{Smith2019} notes that the photometric measurement error in \cite{Aadland2018} is bigger than, for instance, the color difference between O-type and early B-type stars that equals ${\Delta(B-V)\approx0.02-0.03}$ or ${\Delta Q\approx0.05}$. Additionally, \cite{Smith2019} mentioned that uncertainties in zero-point measurement reaches  $0.02\,$mag in $B$ and $V$ filters, and $0.03-0.04\,$mag in the $U$ filter. Furthermore, rms scatters of photometric measurement between images of the same stars can amount to $\sigma_U=0.13$, $\sigma_B=0.07$ and $\sigma_V=0.06$ \citep{Zaritsky2002}. Moreover, it was noted that photometric errors of stars with $V<13.9$ can reach up to $0.1\,$mag depending on the filter \citep{Zaritsky2004}.

Indeed, as stellar atmosphere models show, the dependence of $(B-V)$ color on the effective temperature is significantly weakened for B5-B6 supergiants ($T_{\text{eff}}>15000\,$K) and hotter stars \citep{Castelli2003, Fitzpatrick2005}. The ${(U-B)}$ color index is a sensitive temperature indicator for OB stars \citep{Castelli2003, Martins2006}, however, the uncertainties in the reddening estimates prevent it from being used to accurately diagnose spectral classes \citep{Massey1989}. At the same time, if combined, above-mentioned errors in the UBV photometry can lead to an unpredictable distortion of the reddening-free index $Q$.

All above-mentioned factors significantly influence the fundamental possibility of sampling the most massive stars. To investigate the impact of photometric errors on the results of the study, we compared the spatial distributions of the selected blue stars with those of evolved massive stars identified using various photometric and spectral data.

At first, we filtered the LGGS data with the photometric criteria $Q<-1.20$, ${(U-B)<-0.5}$ and ${(B-V)<0.2}$ to reduce the impact of photometric errors on the results of the study, as was done by \cite{Aadland2018}. In the subsequent work, we used the simple photometric criteria $(U-B)$ and $(B-V)$ in accordance with spectral classification to obtain the optimal sample of massive blue stars. We assume that the reddening will have a limited effect on the selection results, since the study by \cite{Massey1995b} showed low dispersion of values ${E(B-V)=0.13\pm0.02}$ among several OB associations in the M33 galaxy. Further studies presented identical average values of the reddening \citep{Massey2007, Wang2022}. 

We chose two star samples, differentiated by potential initial masses and ages. The first group comprises blue stars with color indices $({U-B)<-1.00}$, ${(B-V)<-0.20}$ and $M_{\rm V}<-5.0$. Those criteria match the early B-type supergiants and hotter stars with luminosity class I \citep{Fitzgerald1970, Fitzpatrick1988, Ivanov1996, Fitzpatrick2005} with initial masses of  $M_{\rm init}\gtrsim20\,M_\odot$ \citep{Massey1989, Eggenberger2021}. We supplemented the first sample with stars with $Q<-0.96$, which can potentially be the hottest early O-type stars with $\log{T_{\rm eff}} \gtrsim 4.60$, that are more reddened than average blue star in the M33 galaxy. We had also selected O-type stars with $Q<-0.88$, $M_{\rm V}<-5.0$ and sorted them by the distance to nearest ionized hydrogen regions, identified on the H\,II image of the M33 galaxy \citep{Massey2007Halpha} with {\tt SEXTRACTOR} software \citep{Bertin1996}. Cross correlation of lower quartile of the obtained sample with stars matched by $(U-B)$ and $(B-V)$ criteria yielded an additional 85 luminous blue stars, which presumably are highly reddened. We used the value of $Q$-index, corresponding to higher atmospheric temperatures than the color indices presented above, and H\,II data in order to avoid including many lower-mass stars in the sample due to the complex influence of measurement uncertainties on the $Q$ indicator. The second group is represented by blue stars of early spectral types, predominantly of the main sequence with $10\,M_\odot \lesssim M_{\rm init}\lesssim20\,M_\odot$, selected according to relations $({U-B)<-0.90}$, ${(B-V)<-0.25}$ and ${-3.0<M_{\rm V}<-4.0}$ \citep{Fitzgerald1970, Flower1977, Straizys1981, Castelli2003, Fitzpatrick2005}. At the final stage, we eliminated Wolf-Rayet stars from our samples by cross-correlation with \cite{Neugent2011} catalogue. Thus, in total, the first and second groups included 2912 and 5376 stars, respectively. The identification of OB associations in subsequent work was based on the first sample consisting of the most massive blue stars, which are likely to be located near the centers of their parent clusters and are therefore better suited for searching for relatively compact young stellar groups.

\subsection{LBV and LBV candidates sample}

In our study, we used the sample, that including both confirmed LBVs and LBV candidates, similarly to \cite{Smith2015}. Criteria for the inclusion of bright blue stars in the LBV candidate group vary significantly among different studies \citep{Massey2007, Humphreys2017, Valeev2009, Solovyeva2020}. The most strictly defined characteristics of LBVs were given in the work of \cite{Humphreys2017}. In the series of works \cite{Humphreys2014, Humphreys2017}, authors performed a detailed analysis of optical spectra, IR photometry and luminosities of LBV candidates in the M31 and M33 galaxies. The authors divided LBV candidates into several subgroups according to the spectral features: Of/late-WN, Fe\,II-emission line stars, some blue and yellow supergiants.

In current work, we used a list of 4 LBVs and 19 LBV candidates, presented in \cite{Humphreys2017}, which was additionally supplemented with three objects: J013242.26+302114.1, J013312.81+303012.6, and J013332.64+304127.2. The star J013242.26+302114.1 was identified as a B[e]-supergiant (sgB[e]) by \cite{Humphreys2017}, however, \cite{Kraus2019} showed that J013242.26+302114.1 cannot be assigned to this class of objects and determining its evolutionary status requires further research. J013312.81+303012.6 is a new LBV candidate presented in \cite{Martin2023}. \cite{Massey2007} classified the star J013332.64+304127.2 as a hot LBV candidate, whose spectrum has undergone significant changes since the first observations. In 1982, the spectrum showed a broad He\,II~$\lambda4686$ line and N\,III lines specific to late-WN stars \citep{Massey1983}, while another spectrum of the star presented in \cite{Massey2007} demonstrated many Fe\,II and [Fe\,II] lines typical for cooler stars, including LBV/cLBV. The optical spectrum J013332.64+304127.2, obtained in 2010 and used in the selection of LBV candidates by \citep{Humphreys2017}, appeared to be the similar to the spectra of late-WN stars, as was the earlier spectrum observed in 1982. \cite{Humphreys2017} did not include the object J013332.64+304127.2 in the sample of LBV candidates in the M33 galaxy due to the absence of P\,Cyg profiles of the strongest hydrogen emission lines of the Balmer series and He\,I emissions. Thus, due to the uncertainty associated with the significant spectral variability of this object, in the current study, we assume that J013332.64+304127.2 is a LBV candidate. We included the above-mentioned objects in the list of LBV candidates in the M33 galaxy compiled by \cite{Humphreys2017}, so the total number of them was 22 in addition to 4 confirmed LBVs (Var\,B, Var\,C, Var\,2, Var\,83).

\subsection{Clustering algorithms} \label{methods_params}

There is a large number of algorithms for searching for stellar groups \citep{Schmeja2011}. Most often, clustering of star fields is carried out using methods based on searching for the densest areas of points in space. The most notable example of such algorithms is DBSCAN \citep{Prisinzano2022, Dong2023, Gao2023, Alfonso2023}, a close analogue of which has been known in astronomy since the early 90s as the PLC technique \citep{Battinelli1991, Ivanov1996, Borissova2004}. The classic implementation of the DBSCAN method operates with two parameters --- the minimum number of points to form a group $N_{\rm min}$ and the search radius for the nearest node $\epsilon$, which allows searching for clusters on only one spatial scale. Hierarchical clustering is possible using various modified and extended versions of the algorithm, for example, such as HDBSCAN \citep{Campello2013} and OPTICS \citep{Ankerst1999}. OPTICS, similar to DBSCAN, uses $N_{\rm min}$ as one of two input parameters. Additionally, the group registration threshold is determined using a reachability plot for each node, which is an ordered set of distances calculated for the entire data set starting from the selected point until the next one. The maximum ratio of two consecutive distances in the reachability plot is limited to $1-\chi$ where $\chi$ is an arbitrary parameter ranging from 0 to 1. On the other hand, the HDBSCAN algorithm only uses $N_{\rm min}$ to perform optimal data clustering.

In this paper, we used the DBSCAN and OPTICS methods for the identification of OB associations. This choice of algorithms was due to greater flexibility in studying the clustering of stellar samples with different crowding of stars. Such conditions occur in separated parts of the galaxy with different background densities of stars and a different average number of potential clusters per unit area. Note that the result completely depends on the initial parameters of the above-mentioned methods, so the choice of their values must be predetermined and justified in the context of the problem being solved. To obtain the optimal values of $N_{\rm min}$, $\epsilon$ and $N_{\rm min}$, $\chi$ for DBSCAN and OPTICS respectively, we used model distributions of the projected positions of clusters members and background stars. We took into account the properties of the observed environment of LBV/cLBV in the M33 galaxy. The method for generating star fields was based on the Monte Carlo algorithm presented in \cite{Aghakhanloo2017}. Next, we consider the main assumptions adopted for the calculation of these models.

At the initial stage we highlighted circles with a projected radius of 500\,pc centered at 4 LBVs and 22 LBV candidates \citep{Massey2016}. Here we adopt the distance to the M33 galaxy equal to $d=9.2\times10^5$\,pc \citep {Jacobs2009}. Using the coordinates of the brightest blue stars presented in section~\ref{methods_selection} for selected regions, we measured the density of stars and estimated probable number of OB associations by eye. The obtained values were used to simulate random star fields of identical shape and size, assuming a Poisson distribution of the number of objects. Thus, the calculated models contain the features of the stellar environment of LBV/cLBV detected in different parts of the galaxy: in the center, in spiral arms and periphery. After we had modeled the stellar background, number and positions of the clusters centers, we used simple massive star ejection model to calculate the projected locations of groups of OB stars.

Bright blue stars are ejected from the compact centers of clusters consisting of the most massive stars \citep{Gies1987, Gvaramadze2008}. The N-body simulations showed that peak of ejection velocity distribution for massive stars is close to the escape velocity $v_{\rm esc}$ in the central regions of the cluster \citep{Oh2016}. In our simplified model, we assumed that all blue stars are ejected from clusters at one moment in time with a fixed velocity $v_{\rm ej}=10$\,km\,s$^{-1}$, typical for OB stars \citep{Gies1987, Gvaramadze2008, Guo2024}. Directions of stellar motions in the observer plane were equally probable within the angle range from 0 to $2\pi$. For each star, the projected distance from the parent cluster center was sampled from a normal distribution with variance $\sigma=v_{\rm ej}(t_{\rm age} - t_{\rm ej})$, where $t_{\rm age}$ is the randomly chosen age of the cluster, $t_{\rm ej}$ --- age of the cluster at the time of the ejections. The value of $t_{\rm ej}=0.8$\,Myr was chosen based on the results of the modeling of massive stars ejections from young star clusters \citep{Oh2016}.

\cite{Aghakhanloo2017} expressed the ratio of the total number of stars in the cluster to the mass of the cluster $N_* \propto M_{\rm cl}$ and obtained distribution of the number of stars in clusters $\frac{dN_{\rm cl}}{dN_*} \propto N^{-2}_*$ using Schechter function $\frac{dN_{\rm cl}}{dM_{\rm cl}} \propto M_{\rm cl}^{-2}$ \citep{Elmegreen1997}, where $N_{\rm cl}$ is the number of clusters, $N_*$ is the number of the stars in the cluster, $M_{\rm cl}$ is the mass of the cluster. Thus, the total number of cluster members can be sampled from derived distribution:

\begin{equation} \label{eq1}
N_* = \frac{1}{R_*({N_{\rm *max}}^{-1} - {N_{\rm *min}}^{-1}) + {N_{\rm *min}}^{-1}},
\end{equation}

where $N_{\rm *max}$, $N_{\rm *min}$ are the maximum and minimum number of stars in the cluster respectively, $R_*$ is a random number from 0 to 1. Note that estimates of the total number stars $N_*$ included in OB associations depend on the selection criteria and methods for identifying clusters of blue stars. This leads to the fact that boundaries of the range of possible values of $N_*$ given in the literature vary several times \citep{Humphreys1980, Ivanov1991, Massey1995a, Ivanov1996}. The result of an automatic search for OB associations based on photometric data for 8 galaxies showed that the average number of blue stars in the found groups lies within $5\lesssim N_* \lesssim 13$ \citep{Ivanov1996}. Author states that the resulting samples predominantly consist of stars with masses $M\approx20\,M_\odot$. According to equation (14) from \cite{Aghakhanloo2017}, we obtained an approximate range of the number of ejected blue stars $9\lesssim N_* \lesssim 24$ at the initial time $t_{\rm age}=1.3\,$Myr. The lower and upper boundaries were accepted as the values $N_{\rm *min}$ and $N_{ \rm *max}$.

The mass of each star was drawn from the Salpeter initial mass function $\frac{dN_*}{dM} \propto M^{-2.35}$ \citep{Salpeter1955}:

\begin{equation} \label{eq2}
M_* = \left(\frac{1}{R_m({M_{\rm *max}}^{-1.35} - {M_{\rm *min}}^{-1.35}) + {M_{\rm *min}}^{-1.35}}\right)^{0.74},
\end{equation}

where $M_{\rm *max}=120\,M_\odot$, $M_{\rm *min}=20\,M_\odot$ are the maximum and minimum values of the initial mass of the star respectively, $R_m$ -- a random number from 0 to 1. This mass range corresponds to the sample of the brightest blue stars, presented in section~\ref{methods_selection}.

Some stars will have evolved by the considered moment in time corresponding to a randomly chosen cluster age $t_{\rm age}$. To model this process, we used the "main-sequence lifetime\,---\,age"{} relationship, calculated using the BINARY\_C code \citep{Izzard2004, Izzard2006, Izzard2009}. As an upper bound on $t_{\rm age}$, we took the duration of the core hydrogen burning phase for a star with mass $M=20\,M_\odot$, which is about 9.7\,Myr. The lower limit of $t_{\rm age}=1.3\,$Myr was adopted assuming that ejected blue stars must be observed as separate objects with angular resolution $\sim 1.0\, \arcsec$\citep{Massey2006, Massey2016} at a distance of $9.2\times10^5$\,pc \citep{Jacobs2009} at any chosen cluster age within the defined range.

We calculated 1000 star fields with different densities of objects for each set of parameters ($N_{\rm min}$, $\epsilon$) and ($N_{\rm min}$, $\chi$). The minimum possible number of star cluster members $N_{\rm min}$ for both methods varied in the range from 3 to 15. DBSCAN algorithm yields the search radius $\epsilon$ ranged from 1 to 100\,pc. The maximum value of $\epsilon$ in the accepted range was calculated using the relation $\epsilon_{\rm max}\approx1.33 D / \sqrt{N_{\rm min}}$ \citep{Ivanov1996} with the minimum number of stars in the cluster $N_{\rm min}=3$. Additionally, the upper limit was estimated considering the typical sizes of OB associations $D\sim30-130\,$pc \citep{Hodge1985, Efremov1987, Ivanov1987}, which were identified by eye without automatic search methods. We used the full range of possible values of $\chi\in [0,1)$ for OPTICS clustering. The maximum distance between two closest points of one cluster was limited to $\epsilon_{\rm max}=100\,$pc, similar to the upper limit of the range of $\epsilon$ values for the DBSCAN method.

\begin{figure*}[t!]
       \centering
       \includegraphics[width=0.49\linewidth]{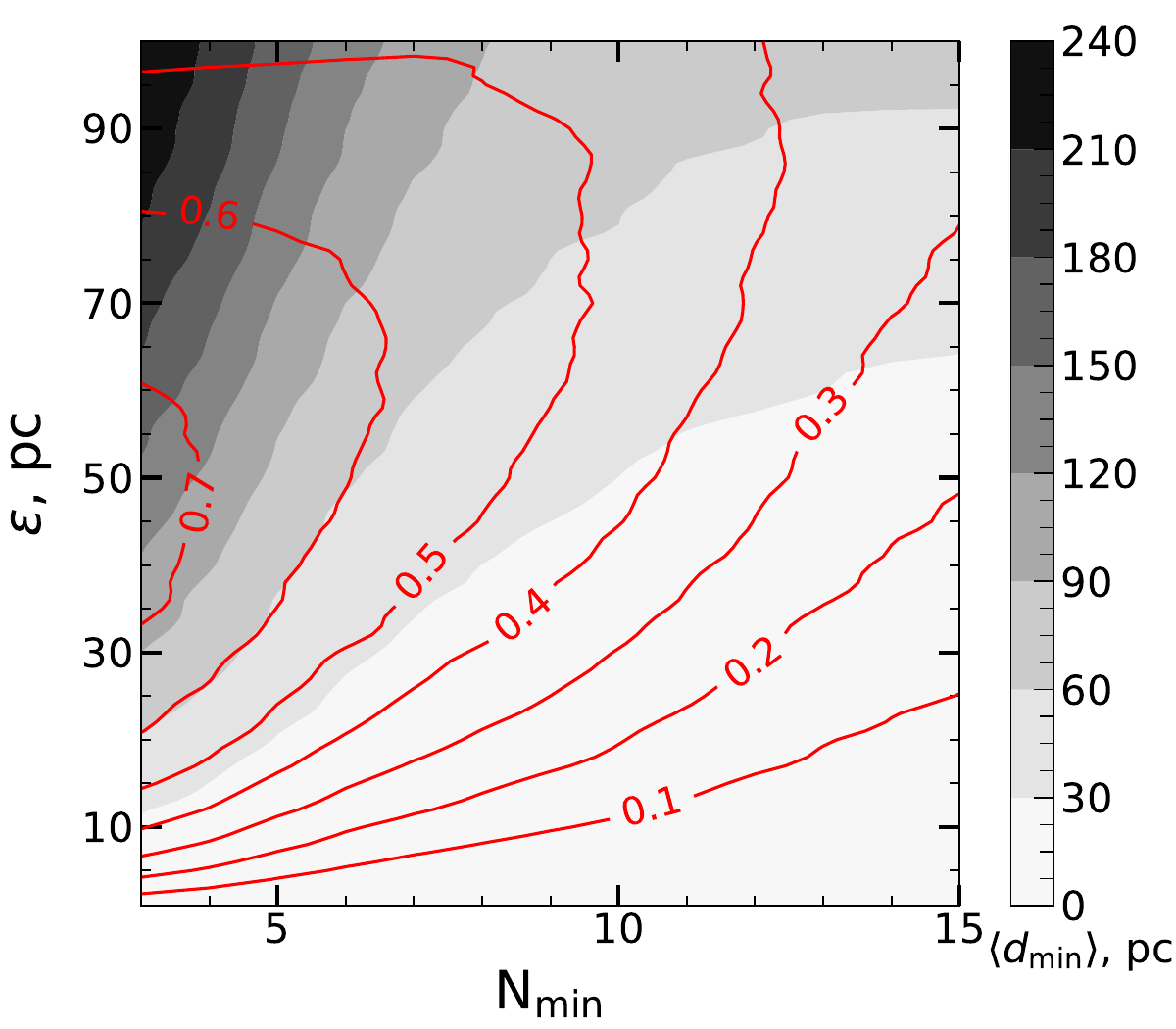}
       \includegraphics[width=0.49\linewidth]{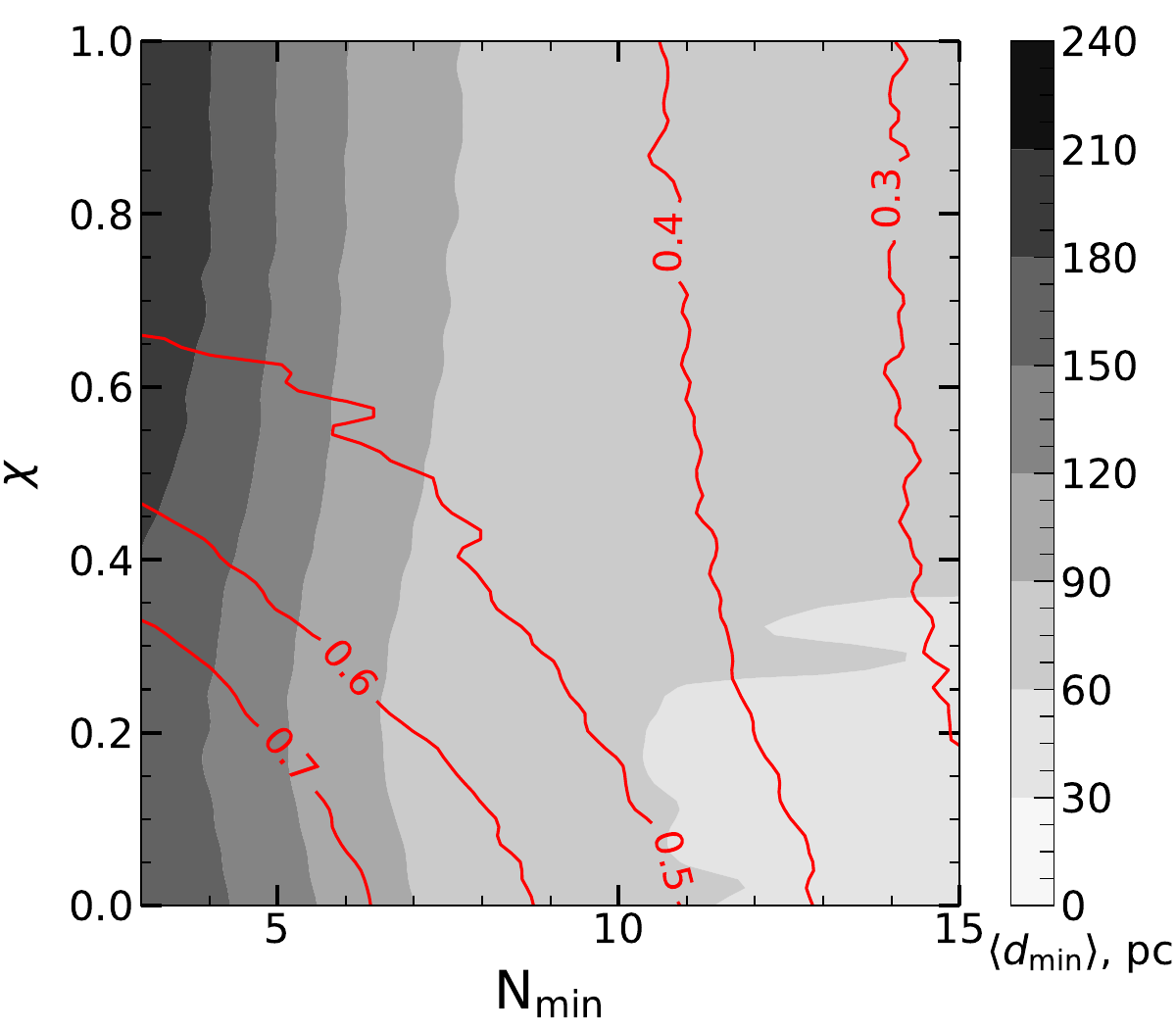}
       \caption{Distribution diagram of $\langle d_{\rm min} \rangle$ (gradient fill) and $\zeta$ (red contour lines) values depending on the initial parameters of the DBSCAN (left) and OPTICS (right) algorithms.}
       \label{fig2}
\end{figure*}

To estimate the performance of DBSCAN and OPTICS on model data, we define two criteria. During the simulation, we measured the distance $d_{\rm min}$ from the center of each found group to the closest initial position of the model cluster. As the first criterion for evaluating the quality of search for star groups, we took the average value $\langle d_{\rm min} \rangle$ among the distances $d_{\rm min}$ obtained from the results of processing all model data. It is expected that the smaller the $\langle d_{\rm min} \rangle$ value, the better the clustering was performed for a given set of parameters. To analyze the quality of the sample, we additionally used the $\zeta$ indicator. This criterion characterizes the percentage of clusters found. The $\zeta$ value was calculated as $\zeta = N_{\rm cl} / N_{\rm init}$, where $N_{\rm cl}$ and $N_{\rm init}$ are the total number of found and generated stellar groups, respectively. A model cluster was considered identified in case its center was located at the smallest distance among all generated clusters to the center of any found stellar group not exceeding typical OB association scale $\sim 80\,$pc \citep{Hodge1985, Efremov1987, Ivanov1987}. Visual inspection of the model data showed that acceptable cluster search results correspond to $\zeta\gtrsim0.7$. Figure~\ref{fig2} demonstrates the dependence of the $\langle d_{\rm min} \rangle$ and $\zeta$ values on the initial parameters of the algorithms.

According to the diagrams shown in the figure~\ref{fig2}, the behavior of $\langle d_{\rm min} \rangle$ and $\zeta$ values near the selected boundary $\zeta=0.7$ is inverse for DBSCAN clustering method. Therefore, adopting the optimal initial parameters for this algorithm, we focused on obtaining the minimum average distance $\langle d_{\rm min} \rangle$ while maintaining the parameter $\zeta$ equal to 0.7. As shown on figure~\ref{fig2}, for the DBSCAN method the value of the search radius $\epsilon$ of projectionally close blue stars lies within the range of 33--50\,pc depending on the chosen value $N_{\rm min}=3$ or $N_{\rm min}=4$. The presented $\epsilon$ range corresponds to the smallest value of $\langle d_{\rm min} \rangle$ at $\zeta\gtrsim0.7$. These estimates of $\epsilon$ are somewhat lower than the values given by other authors. Published works show that the use of algorithms similar to DBSCAN yields an optimal value of the search radius of OB associations of about $\sim60-80$\,pc at $N_{\rm min}=3$ \citep{Battinelli1991, Magnier1993, Ivanov1996}. We assume that such a difference may be due to incompleteness of the observational data. For example, we found approximately 50\% more blue massive stars compared to \cite{Ivanov1996} with similar selection criteria and lower extinction value.

There is no significant (not exceeding the lower bound of the estimates of the characteristic scales of OB associations $\sim30\,$pc) absolute difference between the average distance to the nearest cluster $\Delta \langle d_{\rm min} \rangle\approx30\,$pc for the found ranges of ($N_{\rm min}$, $\epsilon$) values. For that reason at the final stage we used an additional criterion for selecting the initial clustering parameters. The choice of the final ($N_{\rm min}$, $\epsilon$) values among the pairs (3, 33) and (4, 50), corresponding to the minimum $\langle d_{\rm min} \rangle$ at $\zeta=0.7$ and the chosen $N_{\rm min}$, was carried out using an estimate of the average normalized density fluctuation of selected clusters relative to the stellar background \citep{Ivanov1996}:

\begin{equation}
F(\delta)=\left[1/N \sum_{i=1}^{N}(\delta_i - \langle \delta \rangle)^2\right]^{1/2} / \sigma,
\end{equation}

where $\delta_i$ is the individual spatial stellar density in each found OB association, $\langle \delta \rangle$ is the total projected density of blue stars, $\sigma$ is the density dispersion among all selected stellar groups, $N$ is the number of identified OB associations. The value $F(\delta)$ characterizes the dispersion of the stellar densities of the selected groups relative to the stellar background in units of the density dispersion of all found clusters. The maximum value of $F(\delta)$ corresponds to the initial clustering parameters at which OB associations with the highest spatial stellar density (relative to the background) are identified. However, we note that the choice of initial clustering parameters from the entire possible range of values based only on the average normalized density fluctuation $F(\delta)$ gives unsatisfactory results. The maximum $F(\delta)$ is achieved with selecting a few of the densest star-forming regions. For this reason we had used the $F(\delta)$ criterion only at the final stage of choosing initial parameters. The maximum $F(\delta)$ among listed above ($N_{\rm min}$, $\epsilon$) values is reached at $N_{\rm min}=4$ and $\epsilon=50$\,pc for the DBSCAN algorithm.

According to the results of model data clustering using the OPTICS method, the initial parameters ($N_{\rm min}$, $\chi$) values that show the level of completeness of the sample of identified clusters $\zeta\gtrsim0.7$, lie within the ranges of $3 \le N_{\rm min} \le 6$ and $0 \lesssim \chi \lesssim 0.35$. We note that the smaller the minimum number of stars in the cluster $N_{\rm min}$, the wider the boundaries of the permissible values of $\chi$. As shown in figure\,\ref{fig2}, the choice of the optimal value of $\chi$ for a fixed value of $N_{\rm min}$ is difficult, since the average distance to the nearest cluster $\langle d_{\rm min} \rangle$ for $\zeta\gtrsim0.7$ for each considered $N_{\rm min} \le 6$ is practically independent of $\chi$. The smallest $\langle d_{\rm min} \rangle$ values are achieved at $0 \le \chi \lesssim 0.07$ and $ N_{\rm min}=6$. As in the case of DBSCAN, the difference between the best solutions with $ N_{\rm min}=5$ and $ N_{\rm min}=6$ (in terms of $\langle d_{\rm min} \rangle$ indicator) turned out to be insignificant and limited to less than 30\,pc. There is also an uncertainty in the choice of $\chi$ parameter for each selected $N_{\rm min}$. For that reasons we compared the normalized density fluctuation $F(\delta)$ for pairs of ($N_{\rm min}$, $\chi$) values from the ranges presented above, similar to the selection of the initial parameters of the DBSCAN algorithm. As a result, the final initial parameters of the OPTICS clustering were $N_{\rm min}=5$ and $\chi=0.18$.

Summarizing, to study the spatial distribution of massive stars in the M33 galaxy we chose DBSCAN and OPTICS algorithms with initial parameters $N_{\rm min}=4$, $\epsilon=50$\,pc and $N_{\rm min}=5$, $\chi=0.18$, respectively. The simulation results showed that the best outcome in selecting pairs of ($N_{\rm min}$, $\epsilon$) and ($N_{\rm min}$, $\chi$) values can be achieved only using the additional statistical criterion $F(\delta)$, calculated on the basis of observational data. This necessity is caused by the low absolute difference of the obtained $\langle d_{\rm min} \rangle$ in comparison with the characteristic size of OB associations for the regions of the diagrams with $\zeta\gtrsim0.7$.

\section{Results}
\label{results}

\begin{figure*}[t!]
       \centering
       \includegraphics[width=0.49\linewidth]{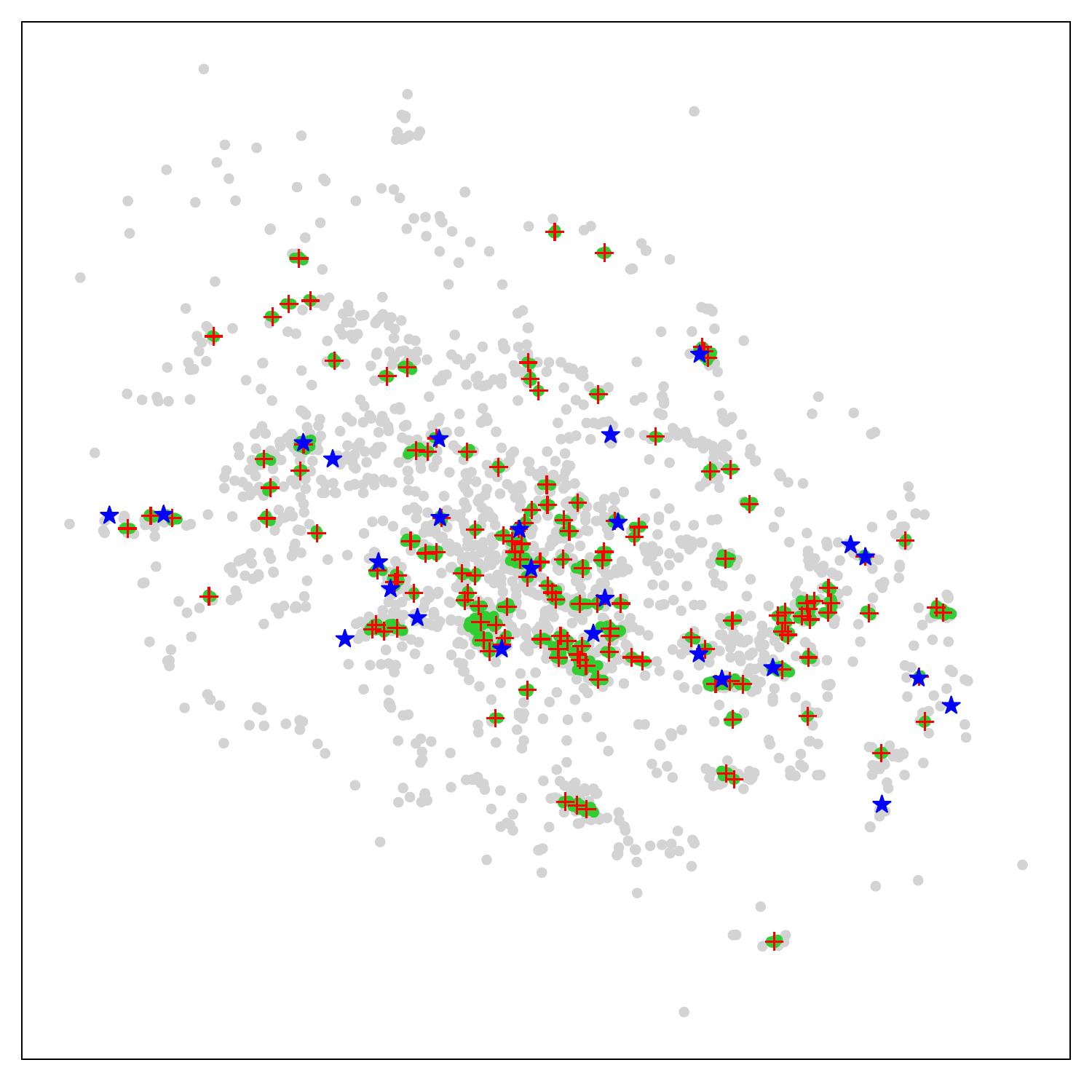}
       \includegraphics[width=0.49\linewidth]{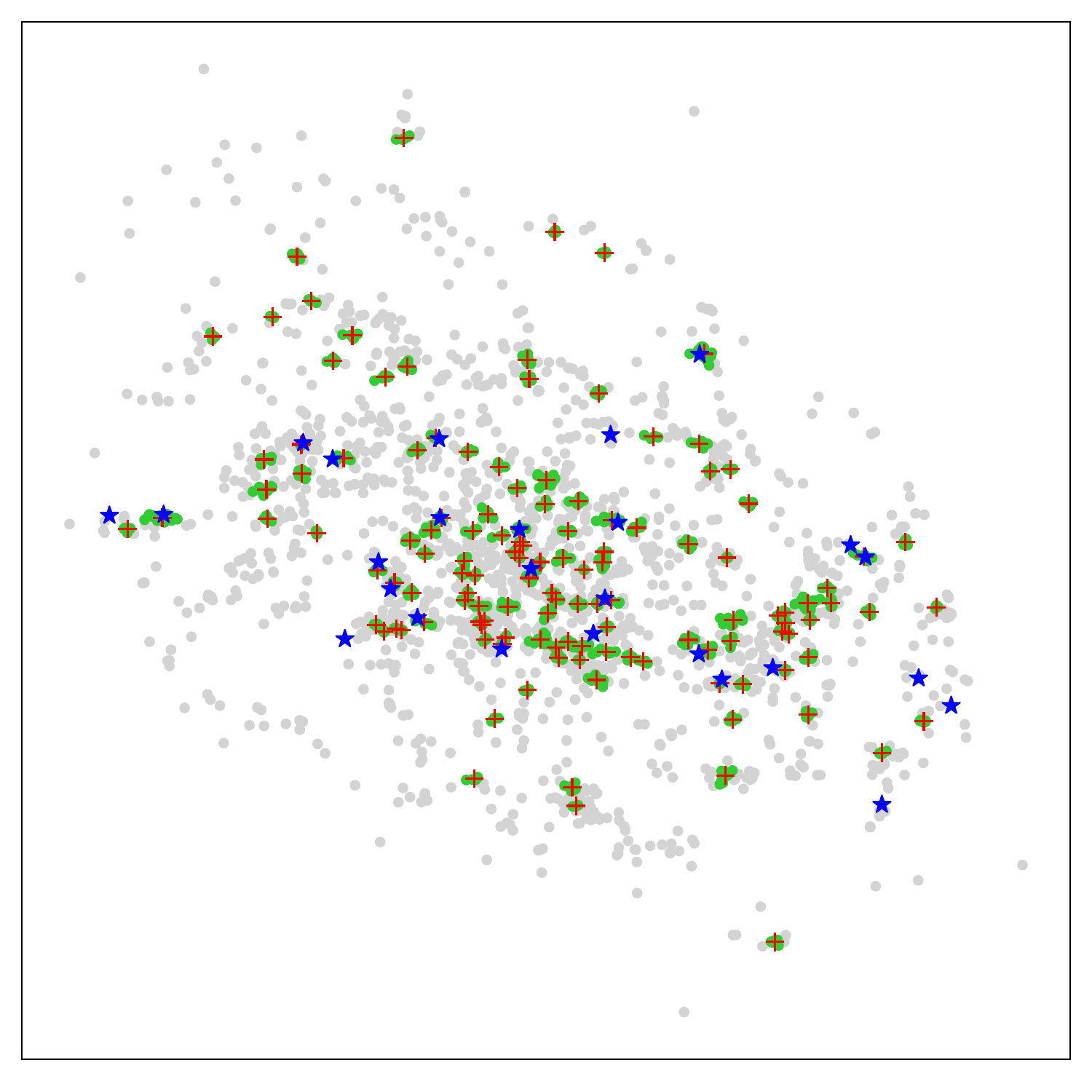}
       \caption{The results of the spatial clustering of a sample of hot massive stars using DBSCAN (left) and OPTICS (right) algorithms. The green dots correspond to the clusters members, the background objects are grey dots. The mass centers of the star groups are indicated by red crosses. LBV and LBV candidates are shown by blue stars.}
       \label{fig3}
\end{figure*}

\begin{figure*}[t!]
       \centering
       \includegraphics[width=0.75\linewidth]{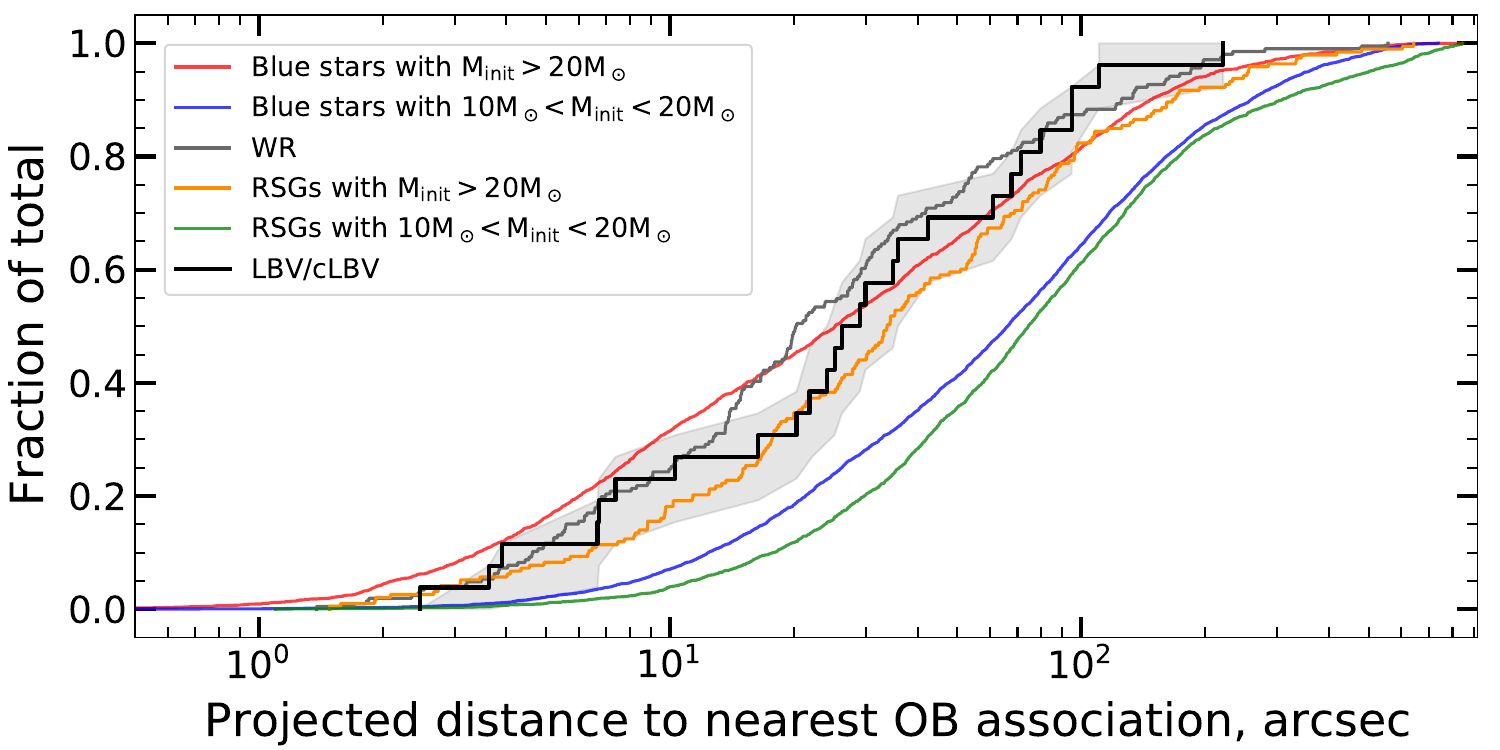}
       \includegraphics[width=0.75\linewidth]{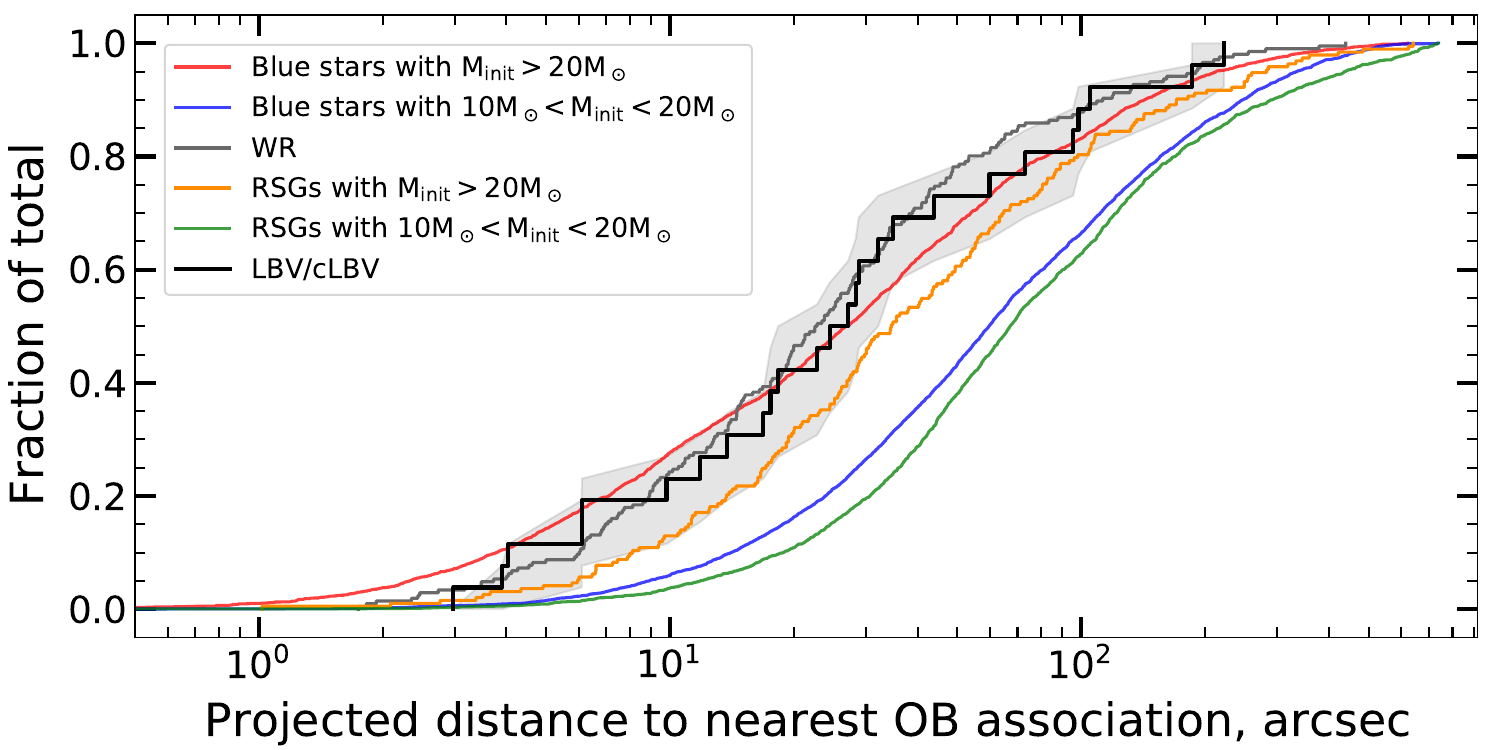}
       \caption{The cumulative distribution functions of the projected distances from stars of different samples to the nearest centers of OB associations, identified by DBSCAN (top) and OPTICS (bottom) clustering algorithms. The 70\% confidence interval for LBV/cLBV distribution is shown by grey color.}
       \label{fig4}
\end{figure*}

\begin{figure*}[t!]
       \centering
       \includegraphics[width=0.75\linewidth]{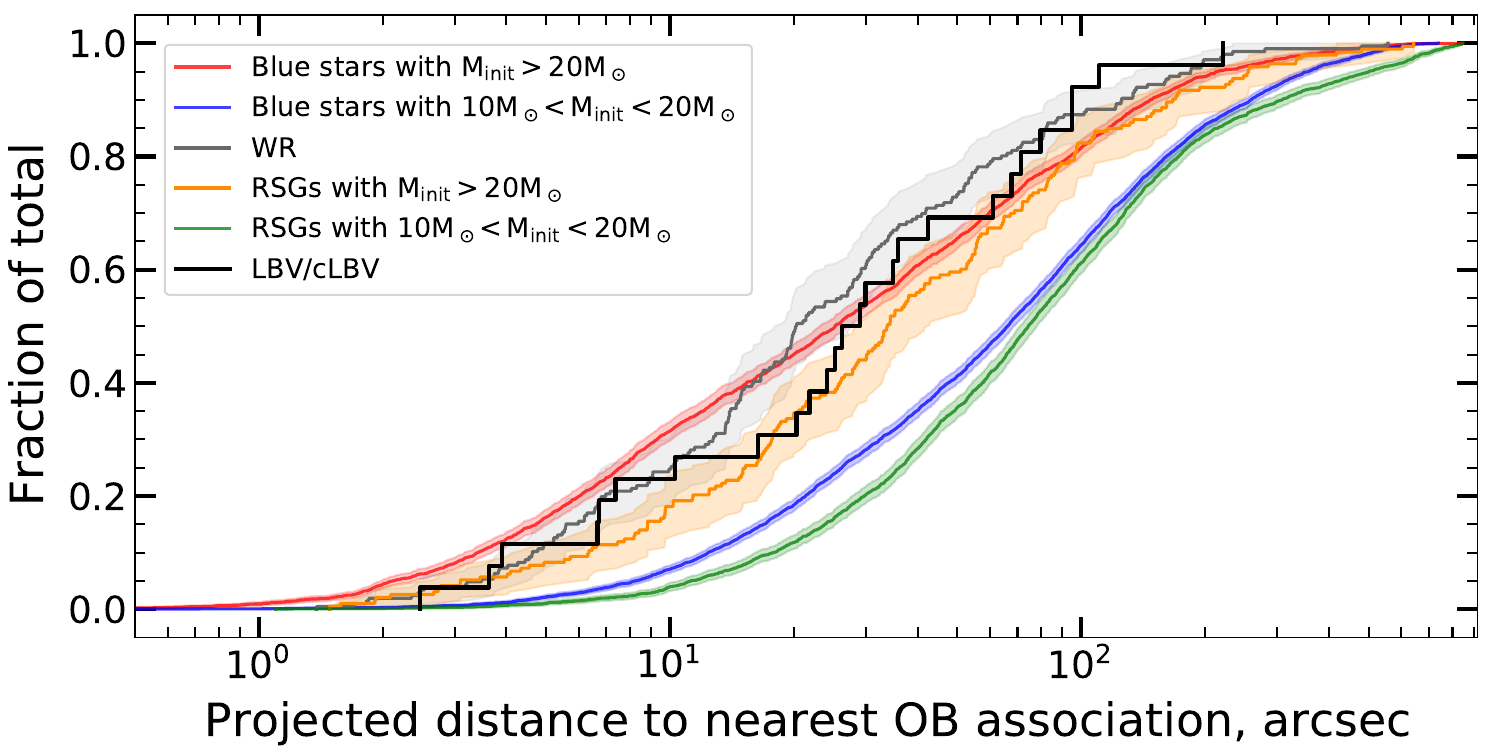}
       \includegraphics[width=0.75\linewidth]{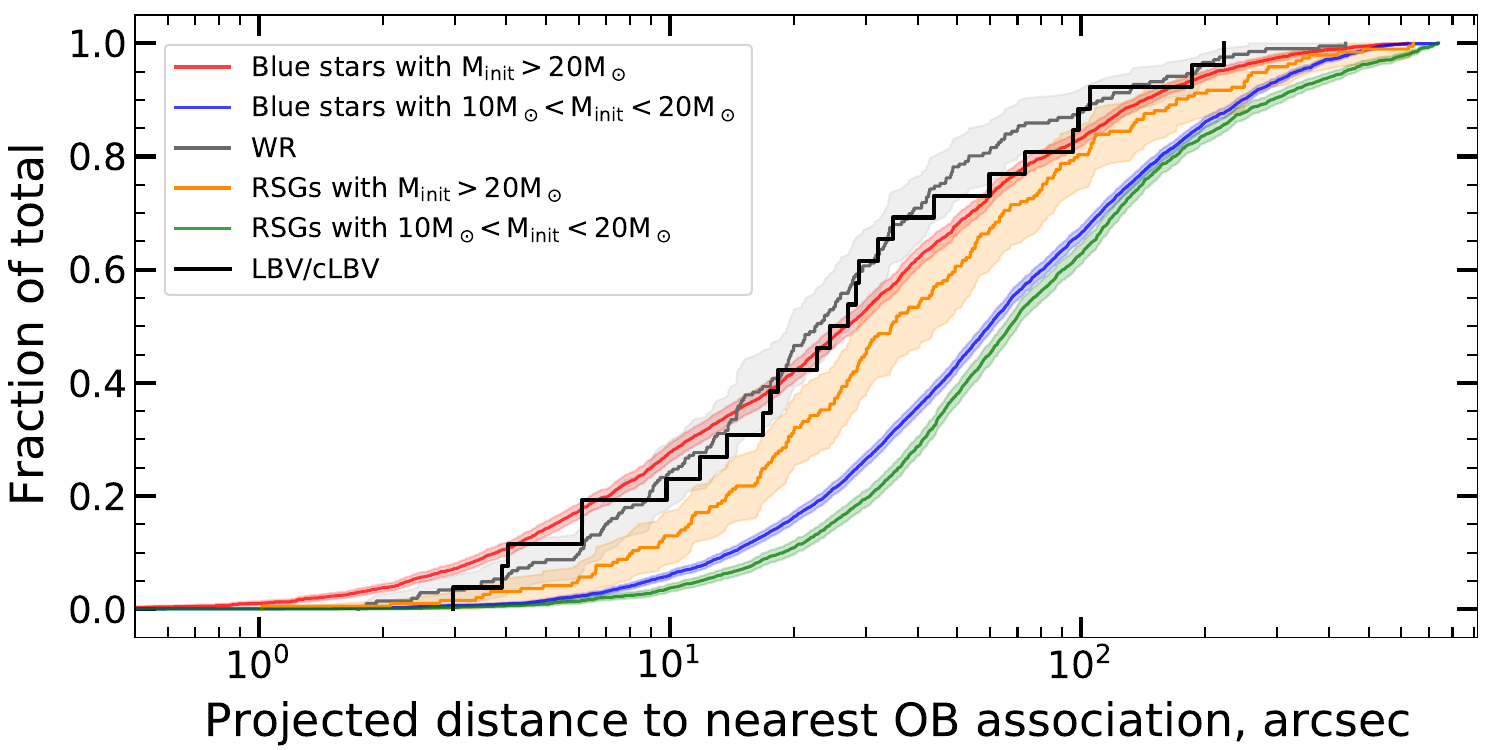}
       \caption{The cumulative distribution functions as in figure~\ref{fig4}. The 95\% confidence intervals for distance distributions of blue stars, WR stars and red supergiants are shown by corresponding colors.}
       \label{fig5}
\end{figure*}

\begin{figure*}[t!]
       \centering
       \includegraphics[width=0.75\linewidth]{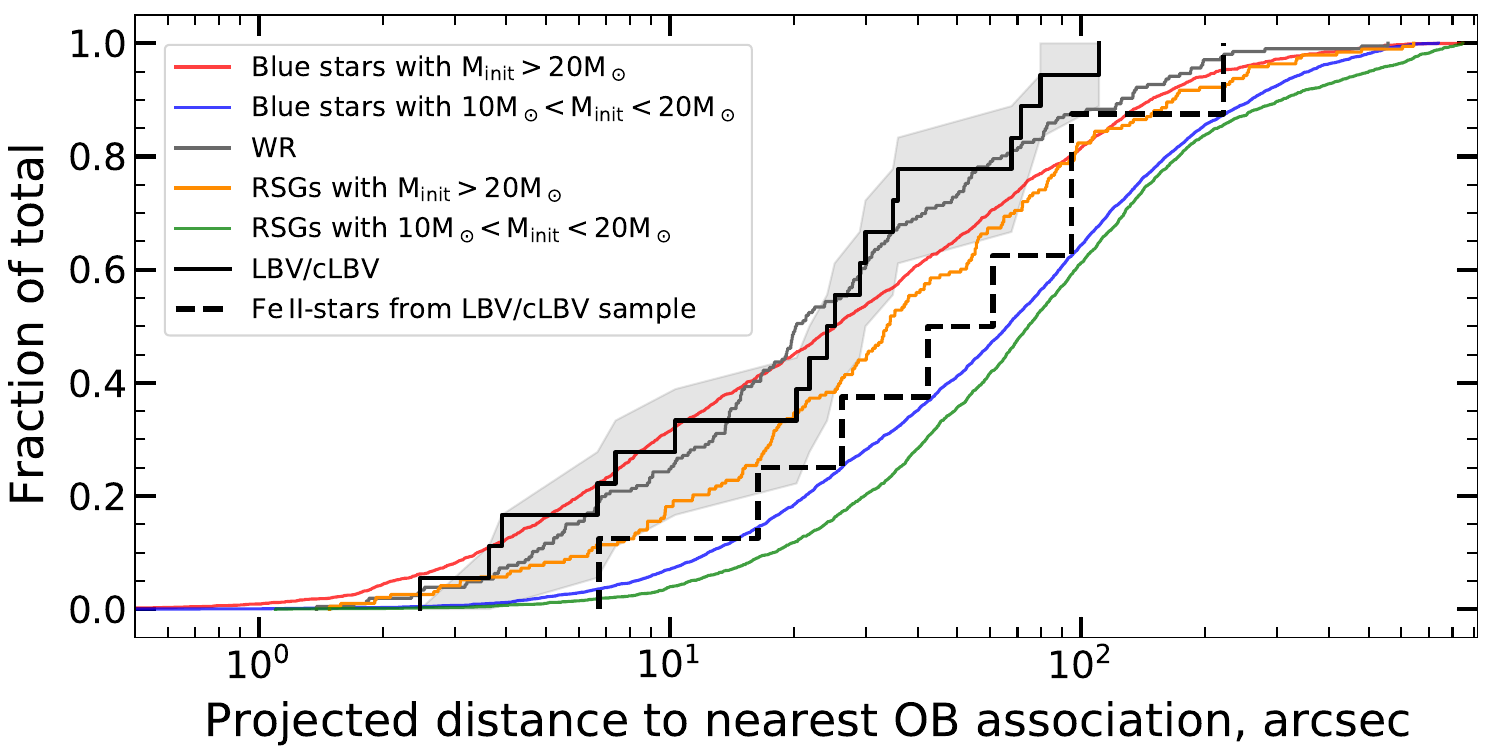}
       \includegraphics[width=0.75\linewidth]{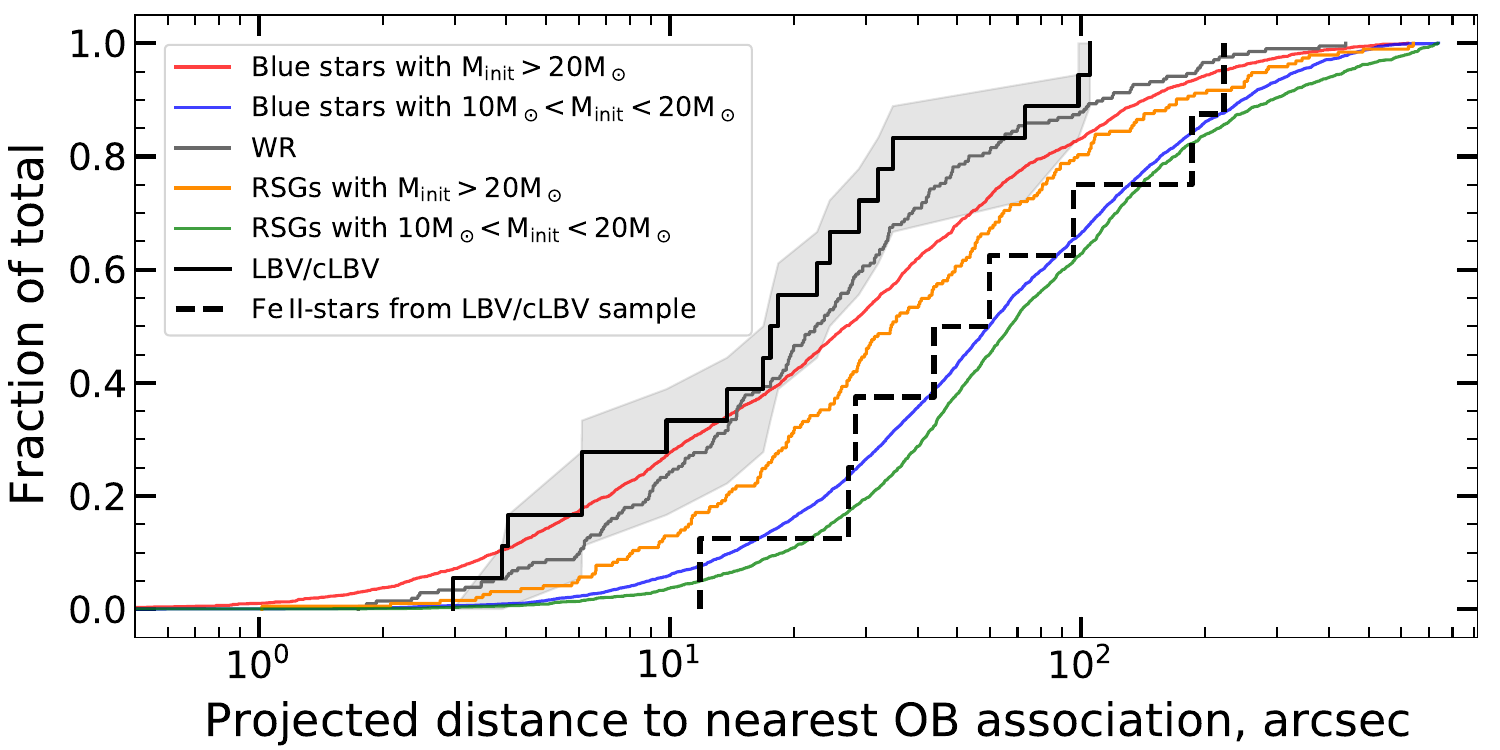}
       \caption{The cumulative distribution functions as in figure~\ref{fig4}, but with Fe-II emission stars separated from the LBV/cLBV sample.}
       \label{fig6}
\end{figure*}

\begin{figure*}[t!]
       \centering
       \includegraphics[width=0.70\linewidth]{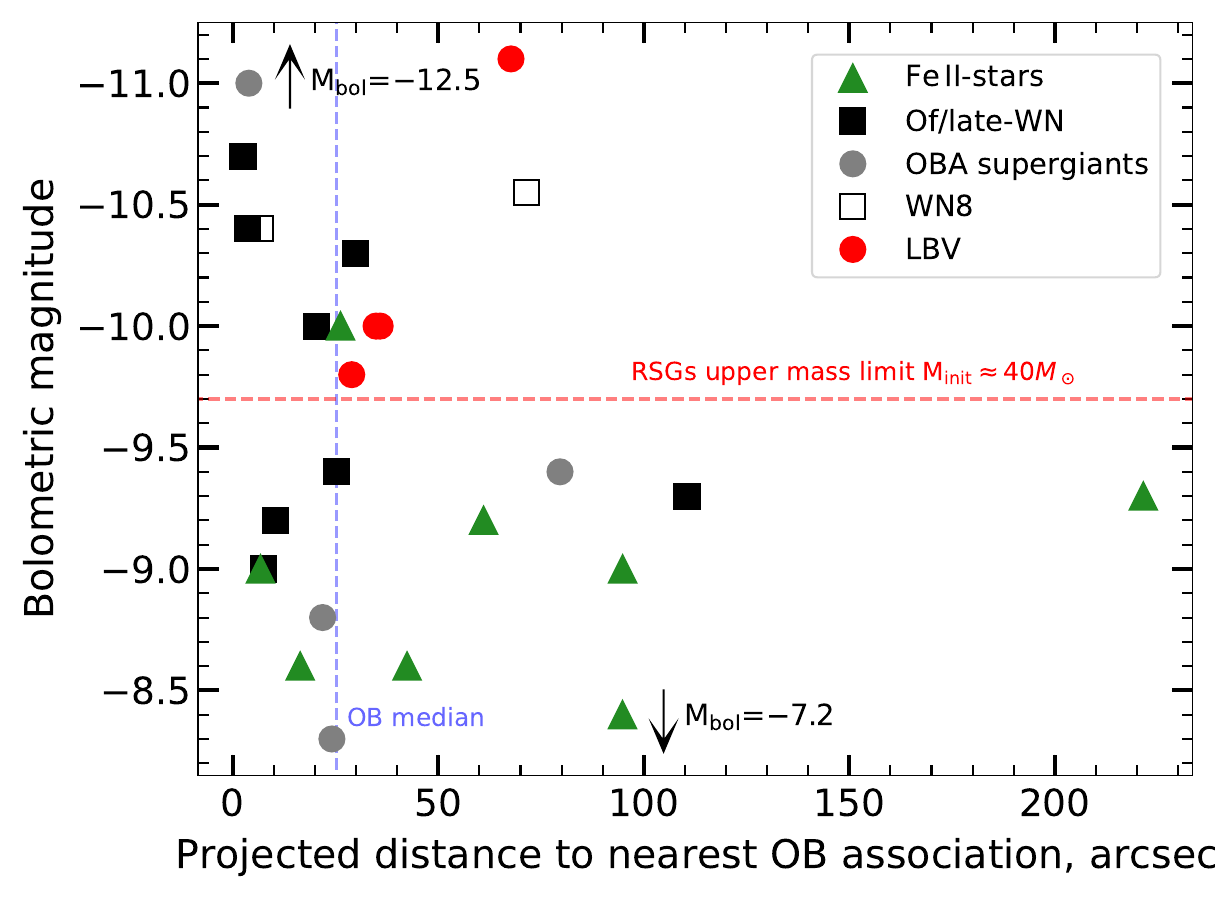}
       \includegraphics[width=0.70\linewidth]{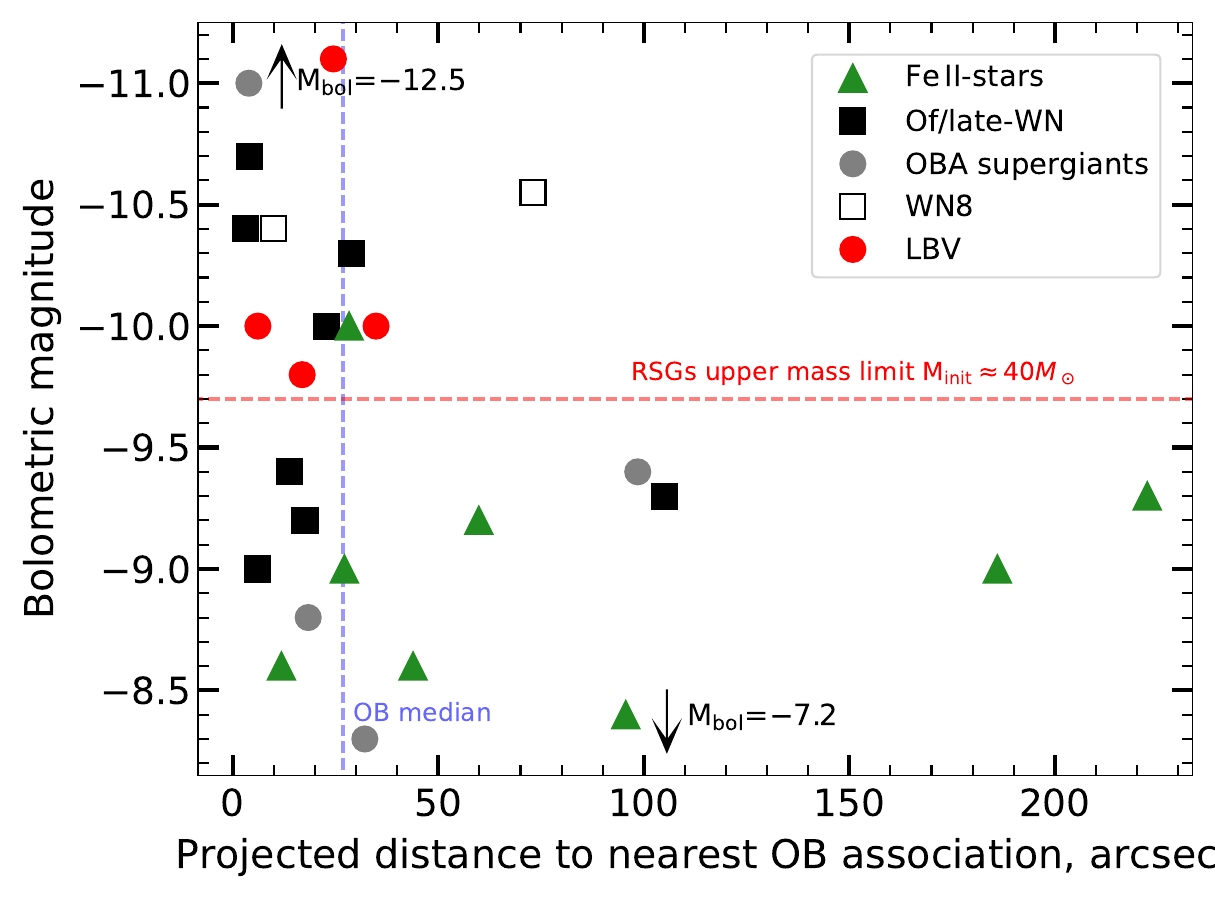}
       \caption{Dependence of the absolute bolometric magnitude on the distance to the nearest center of OB association, identified by DBSCAN (top) and OPTICS (bottom) clustering algorithms, for stars from the LBV/cLBV sample. For comparison, we provide the median value of the distance to the nearest OB association for the sample of massive blue stars (the dashed blue line). The markers of two confirmed LBVs (red circles) are indistinguishable on the top diagram due to the proximity of distances $\sim 35\arcsec$ and identical bolometric magnitudes $M_{\rm bol}=-10\,$ of the corresponding stars.}
       \label{fig7}
\end{figure*} 

In this section we present the main results of searching for the brightest blue stars groups in the M33 galaxy and compare spatial distributions of LBV and some other types of massive stars with respect to the mass centers of the found OB associations.

The figure~\ref{fig3} shows the results of spatial clustering of a sample of 2912 hot stars with $M_{\rm init}>20\,M\odot$, presented in section~\ref{methods_selection}, using the DBSCAN and OPTICS algorithms with the initial parameters defined in section~\ref{methods_params}. As shown in the figure~\ref{fig3}, both methods performed relatively well in finding the most compact blue star groups, and detailed comparison of results showed high similarity of cluster samples found using DBSCAN and OPTICS. However, in extended dense stellar regions, such as in the central part of the galaxy, OPTICS clustering identified as clusters members only a part of the stars from the densest regions of such zones in accordance with the ratio of sequential distances on the reachability plot. Moreover, some low-populated stellar groups consisting of 4 stars that are located mostly in the outer parts of the galaxy were not identified by OPTICS algorithm.

For a comparative analysis of the spatial location of massive stars of different types with respect to the OB associations, we used two stellar samples discussed in section~\ref{methods_selection} and Wolf-Rayet (WR) \citep{Neugent2011}, red supergiant (RSG) \citep{Massey2021} star catalogs of the M33 galaxy. Due to the difference in spatial coverage of the galaxy between observations, we filtered red supergiants that are located further from OB associations than most isolated blue stars ($\sim840\,\arcsec$ or $\sim 3.8\,$kpc). The cumulative distribution functions of the distances from the LBV/cLBV, WR, RSG and OB stars to the centers of the nearest young stellar groups identified by DBSCAN and OPTICS are presented in Fig.~\ref{fig4}. The confidence intervals (CI) for obtained distance distributions are based on the quantiles calculated with bootstrap resampling \citep{Efron1979bootstrap}. On the figure~\ref{fig4}, we follow 70\% CI for LBV/cLBV distance distribution, while figure~\ref{fig5} demonstrates 95\% CI for other distributions.

As shown in figure~\ref{fig4}, DBSCAN algorithm gives a slightly lower median distance value compared to OPTICS among all distributions, which possibly indicates to a better clustering quality. The median distance to the nearest projected center of the OB association for blue massive stars sample is $d_{\rm med}\approx112\,$pc and $d_{\rm med}\approx120\,$pc for DBSCAN and OPTICS, respectively. During data clustering approximately $40\%$ of massive blue stars were marked as members of OB associations. The estimation of the size of identified groups was based on the average distance between the closest members of the association using equations (3) and (4) from \cite{Ivanov1996}. The mean spatial scale of obtained groups is $35-47\,$pc for DBSCAN and $47-63\,$pc for OPTICS depending on the assumption of uniform density within an association. These values are closer to the lower end of typical OB association size estimates $\sim30-130\,$pc presented in the literature \citep{Hodge1985, Efremov1987, Ivanov1987}. Thus, we conclude that our cluster sample contains the most compact and dense blue star groups. The obtained values are comparable with the distance that a star with an initial mass of $20-25\,M\odot$ can move away from a parent cluster with velocity $v_{\rm proj}=10{\rm km}\,{\rm s}^{-1}$ during its main sequence stage. All the estimates listed above depend on the initial parametrization of clusters during modeling, for example, a decrease of initial cluster star count or an increase in ejection velocity result in a higher optimal search distance in the DBSCAN algorithm, which gives a higher cluster detection level at larger scale and, hence, a lower median distance among the sample of blue massive stars and larger cluster sizes.

We found that stars with initial masses of $M_{\rm init}>20\,M_\odot$ and Wolf-Rayet stars are located closest to OB associations. The distribution of blue stars with $M_{\rm init}>20\,M_\odot$ is similar to that of WR stars, despite the fact that they are their younger evolutionary predecessors \citep{Meynet2005, Eggenberger2021}, and therefore should show greater spatial crowding relative to the cluster centers. This effect can be explained by the photometric errors, as mentioned in the section~\ref{methods_selection}. The uncertainty of the stellar magnitudes in the UBV filters does not allow us to clearly select the hottest stars, so the sample will be contaminated by "old"{}, evolved B-supergiants with smaller initial masses. At the same time, WR stars were identified using narrow-band photometry and spectroscopy with the sample incompleteness of only about 5\%  \citep{Neugent2011}. We assume that the spatial distribution of WR stars has a significant influence from the short-lived massive stars with $M_{\rm init} \gtrsim 40\,M_\odot$  that avoid the RSG phase \citep{Eggenberger2021}. The obtained result is in good agreement with the work of \cite{Neugent2011}, where it was noted that the majority of the found WR stars are located in OB associations. Since the ejection efficiency of the most massive single stars from the cluster is not high \citep{Oh2016}, it can be expected that such objects will not have enough time to move significantly away from OB associations \citep{Massey1983, Neugent2011}. 

Red supergiants with initial masses $M_{\rm init}>20\,M_\odot$ are slightly more dispersed from groups of young stars compared to blue massive stars or Wolf-Rayet stars. But as shown in figure~\ref{fig5}, 95\% CI for massive RSG distance distribution moderately crosses that of massive blue stars and Wolf-Rayet stars, hence, we cannot state with high confidence, that red supergiants with initial masses $M_{\rm init}>20\,M_\odot$ are more isolated compared to their evolutionary predecessors or successors.

Blue main-sequence stars with lower masses $10 M_\odot < M_{\rm init} < 20\,M_\odot$ are located significantly further from the cluster centers compared to stars considered above. In turn, red supergiants with initial masses of $10 M_\odot < M_{\rm init} < 20\,M_\odot$, as the oldest stars among the studied star samples, are most isolated from OB associations. Such spatial distribution of stars of various samples with respect to the groups of blue massive stars corresponds to the basic principles of the stellar evolution.

The cumulative distribution function (CDF) of the distances to the nearest OB association for LBV/cLBV is clearly separated from the CDFs of stars with $10 M_\odot < M_{\rm init} < 20\,M_\odot$ and appears closer to that of massive RSGs, but, as it can be seen in figure~\ref{fig4}, even 70\% CI for LBV/cLBV distance distribution significantly overlaps CDF curves for massive blue stars and RSGs or Wolf-Rayet stars.
This, combined with the small difference between CDFs of different massive stars groups, makes it difficult to associate the LBV/cLBV sample with a specific type of massive stars.

Summarizing the clustering results, we conclude that the distribution of distances to the nearest OB association obtained for the LBV/cLBV sample is closer to that of massive stars with $M_{\rm init}>20 M_\odot$. However, as can be seen in figure~\ref{fig3}, the spatial distribution of LBV/cLBV is quite heterogeneous: most of the stars are located near the centers of OB associations, but at the same time, many potentially young massive objects are significantly removed from them. A comparison of the distance to the nearest OB association for stars of different subtypes from the LBV/cLBV sample showed that the most isolated objects are stars with bright Fe\,II-emissions (Fe\,II-emission stars, \citealt{Humphreys2014} or "iron stars"{}, \citealt{Clark2012}). The median absolute bolometric magnitude for the considered Fe\,II-emission stars is $M_{\rm bol}=-9.0$ \citep{Humphreys2017}, which corresponds to an initial mass of about $M_{\rm init}\approx30\,M_\odot$ \citep{Eggenberger2021}, while their spatial distribution (black dashed line, figure~\ref{fig6}) resembles that obtained for blue stars or even red supergiants with smaller masses $10 M_\odot < M_{\rm init} < 20\,M_\odot$. As seen in figure~\ref{fig6}, there is a slight difference between two distributions obtained using DBSCAN and OPTICS. Visual inspection of clustering results in the vicinity of Fe\,II-emission stars shows that OPTICS skipped one low-populated group in their environment, identified by DBSCAN, which affected the CDF curve. We assume that DBSCAN gives more correct results in this case. Nevertheless, regardless of the clustering algorithm used, the Fe\,II-emission stars look more isolated than other stars from LBV/cLBV group, however, small sample size does not allow to make any statistically significant conclusions.

As part of a more detailed study of LBV/cLBV, we compared the projected distances to the centers of the nearest OB association with the absolute magnitudes of the sample objects and grouped them into subtypes \citep{Humphreys2017, Martin2023}. The resulting diagrams are shown in figure ~\ref{fig7}. The brightest and most massive stars with $M_{\rm bol}<-9.7$ and $M_{\rm init}>40\,M_\odot$ demonstrate the highest crowding, regardless of subtype. Indeed, these luminosities correspond to young stars with ages less than 5 million years that are not evolving through red supergiant phase \citep{Eggenberger2021}. A similar result was presented in \cite{Humphreys2016}, where authors studied the distributions of distances to the nearest O-star for objects of the LBV/cLBV sample, divided into two subgroups by potential initial masses. 

Three of the four confirmed LBVs (Var\,B, Var\,C, Var\,2) are found near OB associations. However, Var\,83, which is one of the brightest stars \citep{Valeev2009, Humphreys2017}, appears to be slightly more isolated from groups of massive stars. \cite{Humphreys2016} also noted that Var\,83 is located between two stellar associations A101 and A103 \citep{Humphreys1980}. At the same time, authors showed that three stars in its nearest environment have early spectral types.

As can be seen in figure ~\ref{fig7}, LBV candidates of the Of/late-WN subtype are located closest to the centers of bright blue star groups, and only one such star showed a distance greater than the median for a sample of hot stars with $M_{\rm init}>20\,M_\odot$. In turn, Fe\,II-stars, on average, are significantly more isolated than Of/late-WN stars of comparable luminosities. Two massive stars of the WN8 class should be noted separately -- WR123 (UIT303, HD177230) and V532 (GR290, Romano's star). These objects have hotter atmospheres compared to the Of/late-WN subgroup stars, and past observations of these stars have shown their spectral and photometric variability \citep{Chene2011, Maryeva2019}. For now, V532 is in a stable state, and it is assumed that the star has passed the LBV transitional phase and evolved into a Wolf-Rayet star \citep{Maryeva2019}. In studies of both V532 \citep{Maryeva2019} and some other WN8 stars \citep{Moffat1989} were pointed out their spacing separation from the groups of blue stars. Basing on the hypothesis of \cite{Foellmi2002}, \cite{Chene2011} suggested that pulsations and the spatial isolation of a number of WN8 stars may indicate that such objects are the product of a stellar merging event in a close binary system. However, we note that WR123, unlike V532, is located in the extended H\,II region and is closely surrounded by several blue stars. We assume that WR123 may belong to this compact group of blue stars, which was found by both algorithms, as shown in figure~\ref{fig7}.

\section{Discussion}
\label{discussion}

\begin{figure*}[t!]
       \centering
       \includegraphics[width=0.75\linewidth]{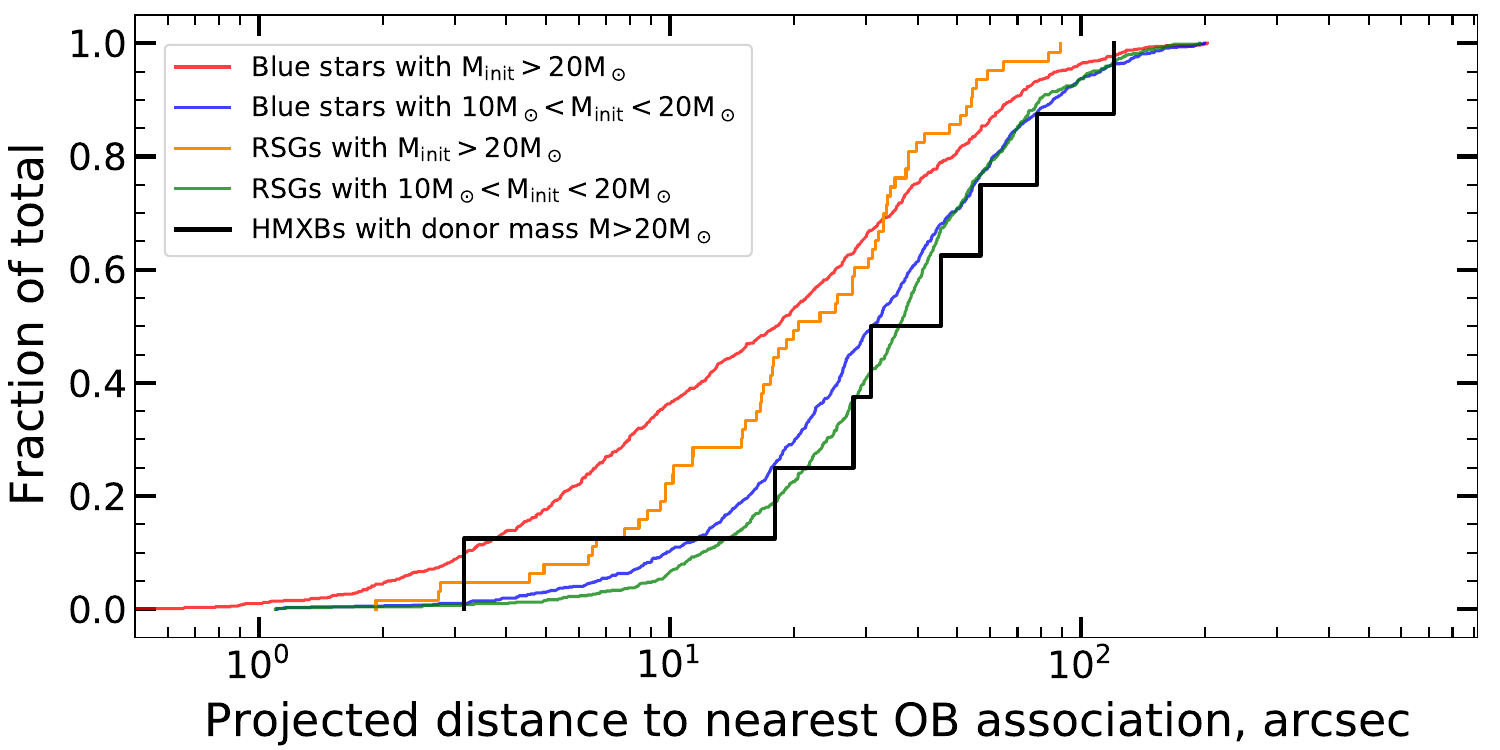}
       \includegraphics[width=0.75\linewidth]{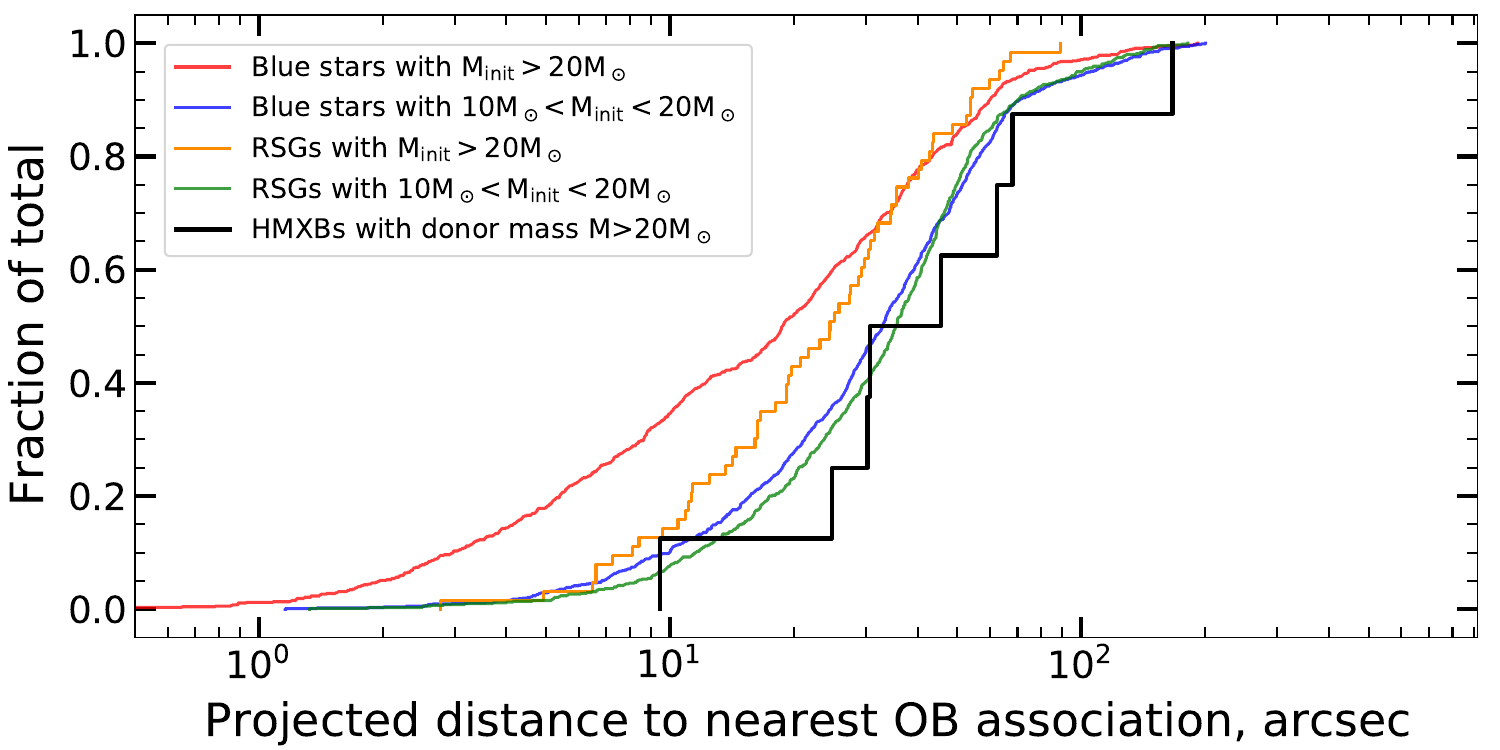}
       \caption{The cumulative distribution functions are similar to those shown in figure~\ref{fig4}, but calculated only for the central part of the M33 galaxy. LBV/cLBV sample was replaced with HMXBs one.}
       \label{fig8}
\end{figure*}

According to the presented results, the spatial distribution of LBV/cLBV with respect to the centers of OB associations is similar to that of the most massive stars with $M_{\rm init}>20\,M_\odot$ and Wolf-Rayet stars. This is consistent with the standard assumption that the nature of luminous blue variables as an intermediate stage in the evolution of single massive stars. Despite the limited LBV/cLBV sample, the spatial distribution of these stars with significant statistical errors differs noticeably from less massive stars with $10 M_\odot < M_{\rm init} < 20\,M_\odot$. The small number of LBV/cLBV could bias the results in the case of non-uniform star distribution, however, as seen in figure~\ref{fig3}, LBV/cLBV are distributed relatively evenly within the galaxy. An additional source of uncertainty in distance estimates is the errors in measuring cluster centers, which is proportional to the dispersion of distances from cluster members to its center averaged around $3-4.3\arcsec$ ($13-19\,$pc) depending on the clustering algorithm. There are also potential error in determining centers and positions of individual stars due to limited angular resolution of observations ($\sim1\arcsec$). We suppose that the value of these errors, which do not exceed the characteristic scale of the selected groups and are significantly smaller than the average distances to the nearest center of OB association for all star samples, could not notably affect the obtained results.

Additionally, it is worth to mention the dependence of the resulting distributions on the initial parameters of the clusters adopted during modeling. The nature of the OB associations is still being investigated and is not fully understood \citep{Wright2020}, and criteria of member selection vary from study to study \citep{Humphreys1980, Hodge1985, Massey1995a}. In fact, OB associations often have a  hierarchical structure, where large stellar groups can be divided to smaller ones and so on, which explains the large variation in star count and size estimates among published works. In our study we decided to use typical cluster star count obtained from previous investigation of OB associations, based on the automatic clustering methods with strict identification criteria \citep{Ivanov1996}. Variation of the initial properties of clusters during modeling can give a different optimal search scale for clustering algorithms and, hence, different size of identified OB associations. We performed clustering with wide parameters ranges for both DBSCAN ($N_{\rm min} \in [3, 5]$, $\epsilon \in [30, 80]\,$pc) and OPTICS ($N_{\rm min} \in [4, 7]$, $\chi \in [0, 0.3]$) to investigate the influence of the initial modeling assumptions to the obtained distributions. As was noted in section~\ref{results}, each set of parameters gives different values of median distance among star samples and mean cluster size, but CDFs curves of LBV/cLBV and other massive stars between adopted parameters look rather similar.

In the end of the paper, we want to discuss objects in the LBV/cLBV sample, particularly Fe\,II-emission stars, which showed significant isolation compared to massive stars with $M_{\rm init}>20\,M_\odot$ or Wolf-Rayet stars. Such discrepancy can be explained by the evolution of Fe\,II-emission stars in massive binary systems, which can be ejected from the clusters at the earliest stages of the cluster evolution due to close encounters \citep{Moeckel2010, Mapelli2011}. To investigate this hypothesis, we compared the spatial distribution of high-mass X-ray binaries (HMXBs) with donor mass $M>20\,M_\odot$ \citep{Lazzarini2023} with stars from some samples presented in the sections~\ref{methods_selection} and \ref{results}. Since the X-ray data are only available for the central regions of the galaxy, we used stars from the corresponding parts of M33. The cumulative distribution functions of projected distances from HMXBs, blue stars and red supergiants to the centers of nearby OB associations are shown in figure~\ref{fig8}.

The resulting distributions showed that HMXBs with donor mass $M>20\,M_\odot$, which were previously massive close binaries, appeared to be more isolated compared to the sample of blue stars and red supergiants with $M_{\rm init}>20\,M_\odot$. Seven out of eight HMXBs located further than $20-25\arcsec$ from the nearest center of OB association, which is comparable to the median distance obtained for massive blue star sample $\sim20\arcsec$ in the central part of the M33 galaxy. The spatial distribution of these HMXBs was found to be close to that of less massive stars with $10 M_\odot < M_{\rm init} < 20\,M_\odot$. This result is in good agreement with studies of the environment of various HMXBs \citep{Binder2023}, including ultra-luminous X-ray sources \citep{Poutanen2013}, which showed that most of HMXBs are on the periphery of young stellar clusters at distances of up to several hundred parsecs. Thus, the isolation of Fe\,II-emission stars can potentially be explained by evolution in a close binary system. Since we have not found in the literature any known cases of accretion from a companion or any other indicators of the binarity of Fe\,II-emission stars, we can assume that additional acceleration to these stars could have been given by a supernova explosion, followed by the disruption of a binary system.

Additionally, some spectral features of Fe\,II-emission stars also favor their evolution in close stellar pairs. The optical spectrum of such stars contains both Fe\,II lines, typical for the relatively cold atmospheres $T_{\rm eff} \lesssim 12000\,$K \citep{Kostenkov2020, Solovyeva2022}, and bright He\,I emissions, observed in the optical spectra of stars with higher photosphere temperatures $T_{\rm eff} \gtrsim 16000\,$K \citep{Najarro1997}. Moreover, in the spectra of some Fe\,II-emission stars, the He\,I emission lines are wider than the Fe\,II ones, which is difficult to explain in the standard spherically symmetric model of the stellar wind. A two-component optical spectrum can be formed in asymmetric winds of fast rotating stars \citep{Maeder2000, Dwarkadas2002}, for example, B[e]-supergiants \citep{Humphreys2014}, whose rotation speed is probably close to the critical one \citep{Kraus2019}. Nonetheless, Fe\,II stars, despite the similarity of their optical spectra with those of sgB[e]s, do not have such high rotation velocities due to the absence of a decretion disk \citep{Humphreys2017}.

Severe wind asymmetry with a difference in the mass loss-rate in order of magnitude between the polar and equatorial directions can also be observed at rotation velocities significantly lower than critical ones, if the star is close enough to the Eddington limit, especially at photosphere temperatures $\log{T_{\rm eff}} \lesssim 4.30$ \citep{Maeder2000}. In a hypothetical evolutionary scenario, Fe\,II-emission stars could spin up significantly as they gain mass from a companion in a close binary system. Then, they slow down to more moderate rotation speeds as they expand during further evolution, while simultaneously losing mass through the stellar wind, achieving larger values of $\Gamma= \frac{\kappa_{\rm es} L}{4 \pi cGM}$. A similar evolutionary path for massive stars in a binary system was suggested by \cite{Smith2015} to describe a possible scenario for the evolution of LBVs that explains their observed projected distribution. As noted by \cite{Smith2015}, the luminosity and current mass of the less massive companion after the mass transfer will not correspond to its age, determined using evolutionary models for single stars. This evolutionary scenario can explain both the isolation of Fe\,II-emission stars and some features of their observed spectrum.

\section{Conclusions}

Our study shows that stellar environment of the confirmed LBVs and LBV candidates of Of/late-WN type in the M33 galaxy, in contrast to Fe\,II-emission stars, are similar to those of single massive stars. This result and arguments, mentioned in section~\ref{discussion}, favors that the origin of Fe\,II-emission stars and other members of LBV/cLBV sample are probably radically different, and Fe\,II-emission stars can be a product of the evolution in a close binary system. Though the lack of statistics on LBV/cLBV stars within a single galaxy prevents us from revealing definitive patterns of their spatial distribution and, hence, clarify their evolutionary status, some of the findings for individual objects is in good agreement with those obtained through a visual examination of their stellar surroundings by other researchers. Further study of projected distribution of LBVs/cLBVs in other galaxies and an increase in the total number of known LBV/cLBV, will allow to make more explicit conclusions about the nature of these objects.

\section*{Acknowledgments}
The work was performed as part of the SAO RAS government contract approved by the Ministry of Science and Higher Education of the Russian Federation. A.E.~Kostenkov is grateful to A.F.~Kholtygin and N.A.~Tikhonov for valuable discussions and thoughtful comments. We thank the anonymous referee for careful reading and constructive remarks.

\bibliographystyle{raa}
\bibliography{ms2024-0343}

@ARTICLE{Conti1975,
       author = {{Conti}, P.~S.},
        title = "{On the relationship between Of and WR stars.}",
      journal = {Memoires of the Societe Royale des Sciences de Liege},
         year = 1975,
        month = jan,
       volume = {9},
        pages = {193-212},
       adsurl = {https://ui.adsabs.harvard.edu/abs/1975MSRSL...9..193C}
}

@INPROCEEDINGS{Conti1984,
       author = {{Conti}, P.~S.},
        title = "{Basic Observational Constraints on the Evolution of Massive Stars}",
    booktitle = {Observational Tests of the Stellar Evolution Theory},
         year = 1984,
       editor = {{Maeder}, A. and {Renzini}, A.},
       volume = {105},
        month = jan,
        pages = {233},
       adsurl = {https://ui.adsabs.harvard.edu/abs/1984IAUS..105..233C}
}

@ARTICLE{Groh2014,
       author = {{Groh}, Jose H. and {Meynet}, Georges and {Ekstr{\"o}m}, Sylvia and {Georgy}, Cyril},
        title = "{The evolution of massive stars and their spectra. I. A non-rotating 60 M$_{{\ensuremath{\odot}}}$ star from the zero-age main sequence to the pre-supernova stage}",
      journal = {\aap},
         year = 2014,
        month = apr,
       volume = {564},
          eid = {A30},
        pages = {A30},
          doi = {10.1051/0004-6361/201322573},
archivePrefix = {arXiv},
       eprint = {1401.7322},
 primaryClass = {astro-ph.SR},
       adsurl = {https://ui.adsabs.harvard.edu/abs/2014A&A...564A..30G}
}

@ARTICLE{Groh2013,
       author = {{Groh}, J.~H. and {Meynet}, G. and {Ekstr{\"o}m}, S.},
        title = "{Massive star evolution: luminous blue variables as unexpected supernova progenitors}",
      journal = {\aap},
         year = 2013,
        month = feb,
       volume = {550},
          eid = {L7},
        pages = {L7},
          doi = {10.1051/0004-6361/201220741},
archivePrefix = {arXiv},
       eprint = {1301.1519},
 primaryClass = {astro-ph.SR},
       adsurl = {https://ui.adsabs.harvard.edu/abs/2013A&A...550L...7G}
}

@ARTICLE{Trundle2008,
       author = {{Trundle}, C. and {Kotak}, R. and {Vink}, J.~S. and {Meikle}, W.~P.~S.},
        title = "{SN 2005 gj: evidence for LBV supernovae progenitors?}",
      journal = {\aap},
         year = 2008,
        month = jun,
       volume = {483},
       number = {3},
        pages = {L47-L50},
          doi = {10.1051/0004-6361:200809755},
archivePrefix = {arXiv},
       eprint = {0804.2392},
 primaryClass = {astro-ph},
       adsurl = {https://ui.adsabs.harvard.edu/abs/2008A&A...483L..47T}
}

@ARTICLE{Andrews2021,
       author = {{Andrews}, Jennifer E. and {Jencson}, Jacob E. and {Van Dyk}, Schuyler D. and {Smith}, Nathan and {Neustadt}, Jack M.~M. and {Sand}, David J. and {Kreckel}, K. and {Kochanek}, C.~S. and {Valenti}, S. and {Strader}, Jay and {Bersten}, M.~C. and {Blanc}, Guillermo A. and {Bostroem}, K. Azalee and {Brink}, Thomas G. and {Emsellem}, Eric and {Filippenko}, Alexei V. and {Folatelli}, Gast{\'o}n and {Kasliwal}, Mansi M. and {Masci}, Frank J. and {McElroy}, Rebecca and {Milisavljevic}, Dan and {Santoro}, Francesco and {Szalai}, Tam{\'a}s},
        title = "{The Blue Supergiant Progenitor of the Supernova Imposter AT 2019krl}",
      journal = {\apj},
         year = 2021,
        month = aug,
       volume = {917},
       number = {2},
          eid = {63},
        pages = {63},
          doi = {10.3847/1538-4357/ac09e1},
archivePrefix = {arXiv},
       eprint = {2009.13541},
 primaryClass = {astro-ph.SR},
       adsurl = {https://ui.adsabs.harvard.edu/abs/2021ApJ...917...63A}
}

@ARTICLE{Smith2016,
       author = {{Smith}, Nathan},
        title = "{The isolation of luminous blue variables: on subdividing the sample}",
      journal = {\mnras},
         year = 2016,
        month = sep,
       volume = {461},
       number = {3},
        pages = {3353-3360},
          doi = {10.1093/mnras/stw1533},
archivePrefix = {arXiv},
       eprint = {1607.01054},
 primaryClass = {astro-ph.SR},
       adsurl = {https://ui.adsabs.harvard.edu/abs/2016MNRAS.461.3353S}
}

@ARTICLE{Aghakhanloo2017,
       author = {{Aghakhanloo}, Mojgan and {Murphy}, Jeremiah W. and {Smith}, Nathan and {Hlo{\v{z}}ek}, Ren{\'e}e},
        title = "{Modelling luminous-blue-variable isolation}",
      journal = {\mnras},
         year = 2017,
        month = nov,
       volume = {472},
       number = {1},
        pages = {591-603},
          doi = {10.1093/mnras/stx2050},
archivePrefix = {arXiv},
       eprint = {1701.05626},
 primaryClass = {astro-ph.SR},
       adsurl = {https://ui.adsabs.harvard.edu/abs/2017MNRAS.472..591A}
}

@ARTICLE{Smith2015,
   author = {{Smith}, N. and {Tombleson}, R.},
    title = "{Luminous blue variables are antisocial: their isolation implies that they are kicked mass gainers in binary evolution}",
  journal = {\mnras},
archivePrefix = "arXiv",
   eprint = {1406.7431},
 primaryClass = "astro-ph.SR",
     year = 2015,
    month = feb,
   volume = 447,
    pages = {598-617},
      doi = {10.1093/mnras/stu2430},
   adsurl = {https://ui.adsabs.harvard.edu/abs/2015MNRAS.447..598S}
}

@ARTICLE{Humphreys2016,
   author = {{Humphreys}, R.~M. and {Weis}, K. and {Davidson}, K. and {Gordon}, M.~S.
	},
    title = "{On the Social Traits of Luminous Blue Variables}",
  journal = {\apj},
archivePrefix = "arXiv",
   eprint = {1603.01278},
 primaryClass = "astro-ph.SR",
     year = 2016,
    month = jul,
   volume = 825,
      eid = {64},
    pages = {64},
      doi = {10.3847/0004-637X/825/1/64},
   adsurl = {https://ui.adsabs.harvard.edu/abs/2016ApJ...825...64H}
}

@ARTICLE{Humphreys1994,
   author = {{Humphreys}, R.~M. and {Davidson}, K.},
    title = "{The luminous blue variables: Astrophysical geysers}",
  journal = {\pasp},
     year = 1994,
    month = oct,
   volume = 106,
    pages = {1025-1051},
      doi = {10.1086/133478},
   adsurl = {https://ui.adsabs.harvard.edu/abs/1994PASP..106.1025H}
}

@ARTICLE{Weis2020,
       author = {{Weis}, Kerstin and {Bomans}, Dominik J.},
        title = "{Luminous Blue Variables}",
      journal = {Galaxies},
         year = 2020,
        month = feb,
       volume = {8},
       number = {1},
        pages = {20},
          doi = {10.3390/galaxies8010020},
archivePrefix = {arXiv},
       eprint = {2009.03144},
 primaryClass = {astro-ph.SR},
       adsurl = {https://ui.adsabs.harvard.edu/abs/2020Galax...8...20W}
}

@ARTICLE{Aadland2018,
       author = {{Aadland}, Erin and {Massey}, Philip and {Neugent}, Kathryn F. and {Drout}, Maria R.},
        title = "{Shedding Light on the Isolation of Luminous Blue Variables}",
      journal = {\aj},
         year = 2018,
        month = dec,
       volume = {156},
       number = {6},
          eid = {294},
        pages = {294},
          doi = {10.3847/1538-3881/aaeb96},
archivePrefix = {arXiv},
       eprint = {1810.11169},
 primaryClass = {astro-ph.SR},
       adsurl = {https://ui.adsabs.harvard.edu/abs/2018AJ....156..294A}
}

@ARTICLE{Smith2019,
       author = {{Smith}, Nathan},
        title = "{The isolation of luminous blue variables resembles aging B-type supergiants, not the most massive unevolved stars}",
      journal = {\mnras},
         year = 2019,
        month = nov,
       volume = {489},
       number = {3},
        pages = {4378-4388},
          doi = {10.1093/mnras/stz2277},
archivePrefix = {arXiv},
       eprint = {1908.06104},
 primaryClass = {astro-ph.SR},
       adsurl = {https://ui.adsabs.harvard.edu/abs/2019MNRAS.489.4378S}
}

@ARTICLE{Kraus2019,
       author = {{Kraus}, Michaela},
        title = "{A Census of B[e] Supergiants}",
      journal = {Galaxies},
         year = 2019,
        month = sep,
       volume = {7},
       number = {4},
        pages = {83},
          doi = {10.3390/galaxies7040083},
archivePrefix = {arXiv},
       eprint = {1909.12199},
 primaryClass = {astro-ph.SR},
       adsurl = {https://ui.adsabs.harvard.edu/abs/2019Galax...7...83K}
}

@ARTICLE{Blaauw1961,
       author = {{Blaauw}, A.},
        title = "{On the origin of the O- and B-type stars with high velocities (the ``run-away'' stars), and some related problems}",
      journal = {\bain},
         year = 1961,
        month = may,
       volume = {15},
        pages = {265},
       adsurl = {https://ui.adsabs.harvard.edu/abs/1961BAN....15..265B}
}

@ARTICLE{Justham2014,
       author = {{Justham}, Stephen and {Podsiadlowski}, Philipp and {Vink}, Jorick S.},
        title = "{Luminous Blue Variables and Superluminous Supernovae from Binary Mergers}",
      journal = {\apj},
         year = 2014,
        month = dec,
       volume = {796},
       number = {2},
          eid = {121},
        pages = {121},
          doi = {10.1088/0004-637X/796/2/121},
archivePrefix = {arXiv},
       eprint = {1410.2426},
 primaryClass = {astro-ph.SR},
       adsurl = {https://ui.adsabs.harvard.edu/abs/2014ApJ...796..121J}
}

@ARTICLE{Gies1987,
       author = {{Gies}, D.~R.},
        title = "{The Kinematical and Binary Properties of Association and Field O Stars}",
      journal = {\apjs},
         year = 1987,
        month = jul,
       volume = {64},
        pages = {545},
          doi = {10.1086/191208},
       adsurl = {https://ui.adsabs.harvard.edu/abs/1987ApJS...64..545G}
}

@ARTICLE{Oh2016,
       author = {{Oh}, Seungkyung and {Kroupa}, Pavel},
        title = "{Dynamical ejections of massive stars from young star clusters under diverse initial conditions}",
      journal = {\aap},
         year = 2016,
        month = may,
       volume = {590},
          eid = {A107},
        pages = {A107},
          doi = {10.1051/0004-6361/201628233},
archivePrefix = {arXiv},
       eprint = {1604.00006},
 primaryClass = {astro-ph.GA},
       adsurl = {https://ui.adsabs.harvard.edu/abs/2016A&A...590A.107O}
}

@ARTICLE{Meynet2005,
       author = {{Meynet}, G. and {Maeder}, A.},
        title = "{Stellar evolution with rotation. XI. Wolf-Rayet star populations at different metallicities}",
      journal = {\aap},
         year = 2005,
        month = jan,
       volume = {429},
        pages = {581-598},
          doi = {10.1051/0004-6361:20047106},
archivePrefix = {arXiv},
       eprint = {astro-ph/0408319},
 primaryClass = {astro-ph},
       adsurl = {https://ui.adsabs.harvard.edu/abs/2005A&A...429..581M}
}

@ARTICLE{Humphreys2017,
       author = {{Humphreys}, Roberta M. and {Gordon}, Michael S. and {Martin}, John C. and {Weis}, Kerstin and {Hahn}, David},
        title = "{Luminous and Variable Stars in M31 and M33. IV. Luminous Blue Variables, Candidate LBVs, B[e] Supergiants, and the Warm Hypergiants: How to Tell Them Apart}",
      journal = {\apj},
         year = 2017,
        month = feb,
       volume = {836},
       number = {1},
          eid = {64},
        pages = {64},
          doi = {10.3847/1538-4357/aa582e},
archivePrefix = {arXiv},
       eprint = {1611.07986},
 primaryClass = {astro-ph.SR},
       adsurl = {https://ui.adsabs.harvard.edu/abs/2017ApJ...836...64H}
}

@ARTICLE{Massey2016,
       author = {{Massey}, Philip and {Neugent}, Kathryn F. and {Smart}, Brianna M.},
        title = "{A Spectroscopic Survey of Massive Stars in M31 and M33}",
      journal = {\aj},
         year = 2016,
        month = sep,
       volume = {152},
       number = {3},
          eid = {62},
        pages = {62},
          doi = {10.3847/0004-6256/152/3/62},
archivePrefix = {arXiv},
       eprint = {1604.00112},
 primaryClass = {astro-ph.SR},
       adsurl = {https://ui.adsabs.harvard.edu/abs/2016AJ....152...62M}
}

@ARTICLE{Massey1995a,
       author = {{Massey}, Philip and {Armandroff}, Taft E. and {Pyke}, Randall and {Patel}, Kanan and {Wilson}, Christine D.},
        title = "{Hot, Luminous Stars in Selected Regions of NGC 6822, M31, and M33}",
      journal = {\aj},
         year = 1995,
        month = dec,
       volume = {110},
        pages = {2715},
          doi = {10.1086/117725},
       adsurl = {https://ui.adsabs.harvard.edu/abs/1995AJ....110.2715M}
}

@ARTICLE{Massey1995b,
       author = {{Massey}, Philip and {Lang}, Cornelia C. and {Degioia-Eastwood}, Kathleen and {Garmany}, Catharine D.},
        title = "{Massive Stars in the Field and Associations of the Magellanic Clouds: The Upper Mass Limit, the Initial Mass Function, and a Critical Test of Main-Sequence Stellar Evolutionary Theory}",
      journal = {\apj},
         year = 1995,
        month = jan,
       volume = {438},
        pages = {188},
          doi = {10.1086/175064},
       adsurl = {https://ui.adsabs.harvard.edu/abs/1995ApJ...438..188M}
}

@ARTICLE{Eggenberger2021,
       author = {{Eggenberger}, Patrick and {Ekstr{\"o}m}, Sylvia and {Georgy}, Cyril and {Martinet}, S{\'e}bastien and {Pezzotti}, Camilla and {Nandal}, Devesh and {Meynet}, Georges and {Buldgen}, Ga{\"e}l and {Salmon}, S{\'e}bastien and {Haemmerl{\'e}}, Lionel and {Maeder}, Andr{\'e} and {Hirschi}, Raphael and {Yusof}, Norhasliza and {Groh}, Jos{\'e} and {Farrell}, Eoin and {Murphy}, Laura and {Choplin}, Arthur},
        title = "{Grids of stellar models with rotation. VI. Models from 0.8 to 120 M$_{{\ensuremath{\odot}}}$ at a metallicity Z = 0.006}",
      journal = {\aap},
         year = 2021,
        month = aug,
       volume = {652},
          eid = {A137},
        pages = {A137},
          doi = {10.1051/0004-6361/202141222},
archivePrefix = {arXiv},
       eprint = {2201.12262},
 primaryClass = {astro-ph.GA},
       adsurl = {https://ui.adsabs.harvard.edu/abs/2021A&A...652A.137E}
}

@ARTICLE{Kang2012,
       author = {{Kang}, Xiaoyu and {Chang}, Ruixiang and {Yin}, Jun and {Hou}, Jinliang and {Zhang}, Fenghui and {Zhang}, Yu and {Han}, Zhanwen},
        title = "{The evolution and star-formation history of M33}",
      journal = {\mnras},
         year = 2012,
        month = oct,
       volume = {426},
       number = {2},
        pages = {1455-1464},
          doi = {10.1111/j.1365-2966.2012.21778.x},
archivePrefix = {arXiv},
       eprint = {1207.5280},
 primaryClass = {astro-ph.CO},
       adsurl = {https://ui.adsabs.harvard.edu/abs/2012MNRAS.426.1455K}
}

@ARTICLE{Massey2005,
       author = {{Massey}, Philip and {Puls}, Joachim and {Pauldrach}, A.~W.~A. and {Bresolin}, Fabio and {Kudritzki}, Rolf P. and {Simon}, Theodore},
        title = "{The Physical Properties and Effective Temperature Scale of O-Type Stars as a Function of Metallicity. II. Analysis of 20 More Magellanic Cloud Stars and Results from the Complete Sample}",
      journal = {\apj},
         year = 2005,
        month = jul,
       volume = {627},
       number = {1},
        pages = {477-519},
          doi = {10.1086/430417},
archivePrefix = {arXiv},
       eprint = {astro-ph/0503464},
 primaryClass = {astro-ph},
       adsurl = {https://ui.adsabs.harvard.edu/abs/2005ApJ...627..477M}
}

@ARTICLE{Zaritsky2002,
       author = {{Zaritsky}, Dennis and {Harris}, Jason and {Thompson}, Ian B. and {Grebel}, Eva K. and {Massey}, Philip},
        title = "{The Magellanic Clouds Photometric Survey: The Small Magellanic Cloud Stellar Catalog and Extinction Map}",
      journal = {\aj},
         year = 2002,
        month = feb,
       volume = {123},
       number = {2},
        pages = {855-872},
          doi = {10.1086/338437},
archivePrefix = {arXiv},
       eprint = {astro-ph/0110665},
 primaryClass = {astro-ph},
       adsurl = {https://ui.adsabs.harvard.edu/abs/2002AJ....123..855Z}
}

@ARTICLE{Zaritsky2004,
       author = {{Zaritsky}, Dennis and {Harris}, Jason and {Thompson}, Ian B. and {Grebel}, Eva K.},
        title = "{The Magellanic Clouds Photometric Survey: The Large Magellanic Cloud Stellar Catalog and Extinction Map}",
      journal = {\aj},
         year = 2004,
        month = oct,
       volume = {128},
       number = {4},
        pages = {1606-1614},
          doi = {10.1086/423910},
archivePrefix = {arXiv},
       eprint = {astro-ph/0407006},
 primaryClass = {astro-ph},
       adsurl = {https://ui.adsabs.harvard.edu/abs/2004AJ....128.1606Z}
}

@ARTICLE{Fitzpatrick2005,
       author = {{Fitzpatrick}, E.~L. and {Massa}, D.},
        title = "{Determining the Physical Properties of the B Stars. II. Calibration of Synthetic Photometry}",
      journal = {\aj},
         year = 2005,
        month = mar,
       volume = {129},
       number = {3},
        pages = {1642-1662},
          doi = {10.1086/427855},
archivePrefix = {arXiv},
       eprint = {astro-ph/0412542},
 primaryClass = {astro-ph},
       adsurl = {https://ui.adsabs.harvard.edu/abs/2005AJ....129.1642F}
}

@INPROCEEDINGS{Castelli2003,
       author = {{Castelli}, F. and {Kurucz}, R.~L.},
        title = "{New Grids of ATLAS9 Model Atmospheres}",
    booktitle = {Modelling of Stellar Atmospheres},
         year = 2003,
       editor = {{Piskunov}, N. and {Weiss}, W.~W. and {Gray}, D.~F.},
       volume = {210},
        month = jan,
        pages = {A20},
          doi = {10.48550/arXiv.astro-ph/0405087},
archivePrefix = {arXiv},
       eprint = {astro-ph/0405087},
 primaryClass = {astro-ph},
       adsurl = {https://ui.adsabs.harvard.edu/abs/2003IAUS..210P.A20C}
}

@ARTICLE{Martins2006,
       author = {{Martins}, F. and {Plez}, B.},
        title = "{UBVJHK synthetic photometry of Galactic O stars}",
      journal = {\aap},
         year = 2006,
        month = oct,
       volume = {457},
       number = {2},
        pages = {637-644},
          doi = {10.1051/0004-6361:20065753},
archivePrefix = {arXiv},
       eprint = {astro-ph/0606587},
 primaryClass = {astro-ph},
       adsurl = {https://ui.adsabs.harvard.edu/abs/2006A&A...457..637M}
}

@ARTICLE{Massey1989,
       author = {{Massey}, Philip and {Garmany}, Catharine D. and {Silkey}, Mariabeth and {Degioia-Eastwood}, Kathleen},
        title = "{The Stellar Content of Two OB Associations in the LMC: LH 117 (NGC 2122) and LH 118}",
      journal = {\aj},
         year = 1989,
        month = jan,
       volume = {97},
        pages = {107},
          doi = {10.1086/114961},
       adsurl = {https://ui.adsabs.harvard.edu/abs/1989AJ.....97..107M}
}

@ARTICLE{Massey2007,
       author = {{Massey}, Philip and {Olsen}, K.~A.~G. and {Hodge}, Paul W. and {Jacoby}, George H. and {McNeill}, Reagin T. and {Smith}, R.~C. and {Strong}, Shay B.},
        title = "{A Survey of Local Group Galaxies Currently Forming Stars. II. UBVRI Photometry of Stars in Seven Dwarfs and a Comparison of the Entire Sample}",
      journal = {\aj},
         year = 2007,
        month = may,
       volume = {133},
       number = {5},
        pages = {2393-2417},
          doi = {10.1086/513319},
archivePrefix = {arXiv},
       eprint = {astro-ph/0702236},
 primaryClass = {astro-ph},
       adsurl = {https://ui.adsabs.harvard.edu/abs/2007AJ....133.2393M}
}

@ARTICLE{Wang2022,
       author = {{Wang}, Yuxi and {Gao}, Jian and {Ren}, Yi and {Chen}, Bingqiu},
        title = "{Dust Extinction Law in Nearby Star-resolved Galaxies. II. M33 Traced by Supergiants}",
      journal = {\apjs},
         year = 2022,
        month = jun,
       volume = {260},
       number = {2},
          eid = {41},
        pages = {41},
          doi = {10.3847/1538-4365/ac63c1},
archivePrefix = {arXiv},
       eprint = {2204.05548},
 primaryClass = {astro-ph.GA},
       adsurl = {https://ui.adsabs.harvard.edu/abs/2022ApJS..260...41W}
}

@ARTICLE{Fitzpatrick1988,
       author = {{Fitzpatrick}, Edward L.},
        title = "{The Properties of OB Supergiants in the Large Magellanic Cloud. II. Spectral Types and Intrinsic Colors}",
      journal = {\apj},
         year = 1988,
        month = dec,
       volume = {335},
        pages = {703},
          doi = {10.1086/166960},
       adsurl = {https://ui.adsabs.harvard.edu/abs/1988ApJ...335..703F}
}

@ARTICLE{Ivanov1996,
       author = {{Ivanov}, G.~R.},
        title = "{OB associations in nearby galaxies.}",
      journal = {\aap},
         year = 1996,
        month = jan,
       volume = {305},
        pages = {708},
       adsurl = {https://ui.adsabs.harvard.edu/abs/1996A&A...305..708I}
}

@ARTICLE{Fitzgerald1970,
       author = {{Fitzgerald}, M. Pim},
        title = "{The Intrinsic Colours of Stars and Two-Colour Reddening Lines}",
      journal = {\aap},
         year = 1970,
        month = feb,
       volume = {4},
        pages = {234},
       adsurl = {https://ui.adsabs.harvard.edu/abs/1970A&A.....4..234F}
}

@ARTICLE{Flower1977,
       author = {{Flower}, P.~J.},
        title = "{Transformations from Theoretical H-R Diagrams to C-M Diagrams}",
      journal = {\aap},
         year = 1977,
        month = jan,
       volume = {54},
        pages = {31},
       adsurl = {https://ui.adsabs.harvard.edu/abs/1977A&A....54...31F}
}

@ARTICLE{Bastian2007,
       author = {{Bastian}, N. and {Ercolano}, B. and {Gieles}, M. and {Rosolowsky}, E. and {Scheepmaker}, R.~A. and {Gutermuth}, R. and {Efremov}, Yu.},
        title = "{Hierarchical star formation in M33: fundamental properties of the star-forming regions}",
      journal = {\mnras},
         year = 2007,
        month = aug,
       volume = {379},
       number = {4},
        pages = {1302-1312},
          doi = {10.1111/j.1365-2966.2007.12064.x},
archivePrefix = {arXiv},
       eprint = {0706.0495},
 primaryClass = {astro-ph},
       adsurl = {https://ui.adsabs.harvard.edu/abs/2007MNRAS.379.1302B}
}

@ARTICLE{Gouliermis2000,
       author = {{Gouliermis}, D. and {Kontizas}, M. and {Korakitis}, R. and {Morgan}, D.~H. and {Kontizas}, E. and {Dapergolas}, A.},
        title = "{OB Stellar Associations in the Large Magellanic Cloud: Identification Method}",
      journal = {\aj},
         year = 2000,
        month = apr,
       volume = {119},
       number = {4},
        pages = {1737-1747},
          doi = {10.1086/301288},
       adsurl = {https://ui.adsabs.harvard.edu/abs/2000AJ....119.1737G}
}

@ARTICLE{Battinelli1991,
       author = {{Battinelli}, P.},
        title = "{A new identification technique for OB associations : OB associations in the Small Magellanic Cloud.}",
      journal = {\aap},
         year = 1991,
        month = apr,
       volume = {244},
        pages = {69},
       adsurl = {https://ui.adsabs.harvard.edu/abs/1991A&A...244...69B}
}

@ARTICLE{Borissova2004,
       author = {{Borissova}, J. and {Kurtev}, R. and {Georgiev}, L. and {Rosado}, M.},
        title = "{A catalogue of OB associations in IC 1613}",
      journal = {\aap},
         year = 2004,
        month = jan,
       volume = {413},
        pages = {889-893},
          doi = {10.1051/0004-6361:20031555},
archivePrefix = {arXiv},
       eprint = {astro-ph/0310282},
 primaryClass = {astro-ph},
       adsurl = {https://ui.adsabs.harvard.edu/abs/2004A&A...413..889B}
}

@ARTICLE{Chemel2022,
       author = {{Chemel}, Alexander A. and {de Grijs}, Richard and {Glushkova}, Elena V. and {Dambis}, Andrey K.},
        title = "{Search for OB associations in Gaia Early Data Release 3}",
      journal = {\mnras},
         year = 2022,
        month = sep,
       volume = {515},
       number = {3},
        pages = {4359-4370},
          doi = {10.1093/mnras/stac1780},
archivePrefix = {arXiv},
       eprint = {2206.12935},
 primaryClass = {astro-ph.GA},
       adsurl = {https://ui.adsabs.harvard.edu/abs/2022MNRAS.515.4359C}
}

@ARTICLE{Schmeja2011,
       author = {{Schmeja}, S.},
        title = "{Identifying star clusters in a field: A comparison of different algorithms}",
      journal = {Astronomische Nachrichten},
         year = 2011,
        month = feb,
       volume = {332},
       number = {2},
        pages = {172},
          doi = {10.1002/asna.201011484},
archivePrefix = {arXiv},
       eprint = {1011.5533},
 primaryClass = {astro-ph.GA},
       adsurl = {https://ui.adsabs.harvard.edu/abs/2011AN....332..172S}
}

@ARTICLE{Prisinzano2022,
       author = {{Prisinzano}, L. and {Damiani}, F. and {Sciortino}, S. and {Flaccomio}, E. and {Guarcello}, M.~G. and {Micela}, G. and {Tognelli}, E. and {Jeffries}, R.~D. and {Alcal{\'a}}, J.~M.},
        title = "{Low-mass young stars in the Milky Way unveiled by DBSCAN and Gaia EDR3: Mapping the star forming regions within 1.5 kpc}",
      journal = {\aap},
         year = 2022,
        month = aug,
       volume = {664},
          eid = {A175},
        pages = {A175},
          doi = {10.1051/0004-6361/202243580},
archivePrefix = {arXiv},
       eprint = {2206.00249},
 primaryClass = {astro-ph.GA},
       adsurl = {https://ui.adsabs.harvard.edu/abs/2022A&A...664A.175P}
}

@ARTICLE{Dong2023,
       author = {{Dong}, Fengqiu Adam and {Crowter}, Kathryn and {Meyers}, Bradley W. and {Pleunis}, Ziggy and {Stairs}, Ingrid and {Tan}, Chia Min and {Yu}, Tinyau Timothy and {Boyle}, Patrick J. and {Cook}, Amanda M. and {Fonseca}, Emmanuel and {Gaensler}, B.~M. and {Good}, Deborah C. and {Kaspi}, Victoria and {McKee}, James W. and {Patel}, Chitrang and {Pearlman}, Aaron B.},
        title = "{The second set of pulsar discoveries by CHIME/FRB/Pulsar: 14 rotating radio transients and 7 pulsars}",
      journal = {\mnras},
         year = 2023,
        month = oct,
       volume = {524},
       number = {4},
        pages = {5132-5147},
          doi = {10.1093/mnras/stad2012},
archivePrefix = {arXiv},
       eprint = {2210.09172},
 primaryClass = {astro-ph.HE},
       adsurl = {https://ui.adsabs.harvard.edu/abs/2023MNRAS.524.5132D}
}

@ARTICLE{Gao2023,
       author = {{Gao}, Xinhua and {Fang}, Dan},
        title = "{A clustering study of the old open cluster Trumpler 19}",
      journal = {\apss},
         year = 2023,
        month = sep,
       volume = {368},
       number = {9},
          eid = {73},
        pages = {73},
          doi = {10.1007/s10509-023-04228-9},
       adsurl = {https://ui.adsabs.harvard.edu/abs/2023Ap&SS.368...73G}
}

@ARTICLE{Alfonso2023,
       author = {{Alfonso}, Jeison and {Garc{\'\i}a-Varela}, Alejandro},
        title = "{A Gaia astrometric view of the open clusters Pleiades, Praesepe, and Blanco 1}",
      journal = {\aap},
         year = 2023,
        month = sep,
       volume = {677},
          eid = {A163},
        pages = {A163},
          doi = {10.1051/0004-6361/202346569},
archivePrefix = {arXiv},
       eprint = {2304.00164},
 primaryClass = {astro-ph.GA},
       adsurl = {https://ui.adsabs.harvard.edu/abs/2023A&A...677A.163A}
}

@article{Ankerst1999,
author = {Ankerst, Mihael and Breunig, Markus M. and Kriegel, Hans-Peter and Sander, J\"{o}rg},
title = {OPTICS: ordering points to identify the clustering structure},
year = {1999},
issue_date = {June 1999},
publisher = {Association for Computing Machinery},
address = {New York, NY, USA},
volume = {28},
number = {2},
issn = {0163-5808},
url = {https://doi.org/10.1145/304181.304187},
doi = {10.1145/304181.304187},
journal = {SIGMOD Rec.},
month = {jun},
pages = {49–60},
numpages = {12}
}

@InProceedings{Campello2013,
author="Campello, Ricardo J. G. B.
and Moulavi, Davoud
and Sander, Joerg",
editor="Pei, Jian
and Tseng, Vincent S.
and Cao, Longbing
and Motoda, Hiroshi
and Xu, Guandong",
title="Density-Based Clustering Based on Hierarchical Density Estimates",
booktitle="Advances in Knowledge Discovery and Data Mining",
year="2013",
publisher="Springer Berlin Heidelberg",
address="Berlin, Heidelberg",
pages="160--172",
isbn="978-3-642-37456-2"
}

@ARTICLE{Jacobs2009,
       author = {{Jacobs}, Bradley A. and {Rizzi}, Luca and {Tully}, R. Brent and {Shaya}, Edward J. and {Makarov}, Dmitry I. and {Makarova}, Lidia},
        title = "{The Extragalactic Distance Database: Color-Magnitude Diagrams}",
      journal = {\aj},
         year = 2009,
        month = aug,
       volume = {138},
       number = {2},
        pages = {332-337},
          doi = {10.1088/0004-6256/138/2/332},
archivePrefix = {arXiv},
       eprint = {0902.3675},
 primaryClass = {astro-ph.CO},
       adsurl = {https://ui.adsabs.harvard.edu/abs/2009AJ....138..332J}
}

@ARTICLE{Salpeter1955,
       author = {{Salpeter}, Edwin E.},
        title = "{The Luminosity Function and Stellar Evolution.}",
      journal = {\apj},
         year = 1955,
        month = jan,
       volume = {121},
        pages = {161},
          doi = {10.1086/145971},
       adsurl = {https://ui.adsabs.harvard.edu/abs/1955ApJ...121..161S}
}

@ARTICLE{Straizys1981,
       author = {{Straizys}, V. and {Kuriliene}, G.},
        title = "{Fundamental Stellar Parameters Derived from the Evolutionary Tracks}",
      journal = {\apss},
         year = 1981,
        month = dec,
       volume = {80},
       number = {2},
        pages = {353-368},
          doi = {10.1007/BF00652936},
       adsurl = {https://ui.adsabs.harvard.edu/abs/1981Ap&SS..80..353S}
}

@ARTICLE{Ivanov1991,
       author = {{Ivanov}, G.~R.},
        title = "{The Stellar Content of Associations and Star Complexes in M33}",
      journal = {\apss},
         year = 1991,
        month = apr,
       volume = {178},
       number = {2},
        pages = {227-249},
          doi = {10.1007/BF00643841},
       adsurl = {https://ui.adsabs.harvard.edu/abs/1991Ap&SS.178..227I}
}

@ARTICLE{Izzard2004,
       author = {{Izzard}, Robert G. and {Tout}, Christopher A. and {Karakas}, Amanda I. and {Pols}, Onno R.},
        title = "{A new synthetic model for asymptotic giant branch stars}",
      journal = {\mnras},
         year = 2004,
        month = may,
       volume = {350},
       number = {2},
        pages = {407-426},
          doi = {10.1111/j.1365-2966.2004.07446.x},
archivePrefix = {arXiv},
       eprint = {astro-ph/0402403},
 primaryClass = {astro-ph},
       adsurl = {https://ui.adsabs.harvard.edu/abs/2004MNRAS.350..407I}
}

@ARTICLE{Izzard2006,
       author = {{Izzard}, R.~G. and {Dray}, L.~M. and {Karakas}, A.~I. and {Lugaro}, M. and {Tout}, C.~A.},
        title = "{Population nucleosynthesis in single and binary stars. I. Model}",
      journal = {\aap},
         year = 2006,
        month = dec,
       volume = {460},
       number = {2},
        pages = {565-572},
          doi = {10.1051/0004-6361:20066129},
       adsurl = {https://ui.adsabs.harvard.edu/abs/2006A&A...460..565I}
}

@ARTICLE{Izzard2009,
       author = {{Izzard}, R.~G. and {Glebbeek}, E. and {Stancliffe}, R.~J. and {Pols}, O.~R.},
        title = "{Population synthesis of binary carbon-enhanced metal-poor stars}",
      journal = {\aap},
         year = 2009,
        month = dec,
       volume = {508},
       number = {3},
        pages = {1359-1374},
          doi = {10.1051/0004-6361/200912827},
archivePrefix = {arXiv},
       eprint = {0910.2158},
 primaryClass = {astro-ph.SR},
       adsurl = {https://ui.adsabs.harvard.edu/abs/2009A&A...508.1359I}
}

@ARTICLE{Magnier1993,
       author = {{Magnier}, Eugene A. and {Battinelli}, Paolo and {Lewin}, Walter H.~G. and {Haiman}, Zoltan and {van Paradijs}, Jan and {Hasinger}, Guenther and {Pietsch}, Wolfgang and {Supper}, Rodrigo and {Truemper}, Joachim},
        title = "{Automated identification of OB associations in M 31.}",
      journal = {\aap},
         year = 1993,
        month = oct,
       volume = {278},
        pages = {36-42},
       adsurl = {https://ui.adsabs.harvard.edu/abs/1993A&A...278...36M}
}

@ARTICLE{Massey2021,
       author = {{Massey}, Philip and {Neugent}, Kathryn F. and {Levesque}, Emily M. and {Drout}, Maria R. and {Courteau}, St{\'e}phane},
        title = "{The Red Supergiant Content of M31 and M33}",
      journal = {\aj},
         year = 2021,
        month = feb,
       volume = {161},
       number = {2},
          eid = {79},
        pages = {79},
          doi = {10.3847/1538-3881/abd01f},
archivePrefix = {arXiv},
       eprint = {2011.13279},
 primaryClass = {astro-ph.SR},
       adsurl = {https://ui.adsabs.harvard.edu/abs/2021AJ....161...79M}
}

@ARTICLE{Neugent2011,
       author = {{Neugent}, Kathryn F. and {Massey}, Philip},
        title = "{The Wolf-Rayet Content of M33}",
      journal = {\apj},
         year = 2011,
        month = jun,
       volume = {733},
       number = {2},
          eid = {123},
        pages = {123},
          doi = {10.1088/0004-637X/733/2/123},
archivePrefix = {arXiv},
       eprint = {1103.5549},
 primaryClass = {astro-ph.CO},
       adsurl = {https://ui.adsabs.harvard.edu/abs/2011ApJ...733..123N}
}

@ARTICLE{Massey1983,
       author = {{Massey}, P. and {Conti}, P.~S.},
        title = "{Wolf-rayet stars in M 33.}",
      journal = {\apj},
         year = 1983,
        month = oct,
       volume = {273},
        pages = {576-589},
          doi = {10.1086/161393},
       adsurl = {https://ui.adsabs.harvard.edu/abs/1983ApJ...273..576M}
}

@ARTICLE{Solovyeva2020,
       author = {{Solovyeva}, Y. and {Vinokurov}, A. and {Sarkisyan}, A. and {Atapin}, K. and {Fabrika}, S. and {Valeev}, A.~F. and {Kniazev}, A. and {Sholukhova}, O. and {Maslennikova}, O.},
        title = "{New luminous blue variable candidates in the NGC 247 galaxy}",
      journal = {\mnras},
         year = 2020,
        month = oct,
       volume = {497},
       number = {4},
        pages = {4834-4842},
          doi = {10.1093/mnras/staa2117},
archivePrefix = {arXiv},
       eprint = {2008.06215},
 primaryClass = {astro-ph.SR},
       adsurl = {https://ui.adsabs.harvard.edu/abs/2020MNRAS.497.4834S}
}

@ARTICLE{Humphreys2014,
       author = {{Humphreys}, Roberta M. and {Weis}, Kerstin and {Davidson}, Kris and {Bomans}, D.~J. and {Burggraf}, Birgitta},
        title = "{Luminous and Variable Stars in M31 and M33. II. Luminous Blue Variables, Candidate LBVs, Fe II Emission Line Stars, and Other Supergiants}",
      journal = {\apj},
         year = 2014,
        month = jul,
       volume = {790},
       number = {1},
          eid = {48},
        pages = {48},
          doi = {10.1088/0004-637X/790/1/48},
archivePrefix = {arXiv},
       eprint = {1407.2259},
 primaryClass = {astro-ph.SR},
       adsurl = {https://ui.adsabs.harvard.edu/abs/2014ApJ...790...48H}
}

@ARTICLE{Martin2023,
       author = {{Martin}, John C. and {Humphreys}, Roberta M. and {Weis}, Kerstin and {Bomans}, Dominik J.},
        title = "{A New LBV Candidate in M33}",
      journal = {Research Notes of the American Astronomical Society},
         year = 2023,
        month = may,
       volume = {7},
       number = {5},
          eid = {96},
        pages = {96},
          doi = {10.3847/2515-5172/acd54d},
archivePrefix = {arXiv},
       eprint = {2305.11687},
 primaryClass = {astro-ph.SR},
       adsurl = {https://ui.adsabs.harvard.edu/abs/2023RNAAS...7...96M}
}

@ARTICLE{Efremov1987,
       author = {{Efremov}, Iu. N. and {Ivanov}, G.~R. and {Nikolov}, N.~S.},
        title = "{Star Complexes and Associations in the Andromeda Galaxy}",
      journal = {\apss},
         year = 1987,
        month = jul,
       volume = {135},
       number = {1},
        pages = {119-130},
          doi = {10.1007/BF00644467},
       adsurl = {https://ui.adsabs.harvard.edu/abs/1987Ap&SS.135..119E}
}

@ARTICLE{Hodge1985,
       author = {{Hodge}, P.},
        title = "{The stellar associations of the Small Magellanic Cloud.}",
      journal = {\pasp},
         year = 1985,
        month = jun,
       volume = {97},
        pages = {530-536},
          doi = {10.1086/131564},
       adsurl = {https://ui.adsabs.harvard.edu/abs/1985PASP...97..530H}
}

@ARTICLE{Ivanov1987,
       author = {{Ivanov}, G.~R.},
        title = "{Stellar Associations and Complexes in M33}",
      journal = {\apss},
         year = 1987,
        month = aug,
       volume = {136},
       number = {1},
        pages = {113-128},
          doi = {10.1007/BF00661259},
       adsurl = {https://ui.adsabs.harvard.edu/abs/1987Ap&SS.136..113I}
}

@ARTICLE{Clark2012,
       author = {{Clark}, J.~S. and {Castro}, N. and {Garcia}, M. and {Herrero}, A. and {Najarro}, F. and {Negueruela}, I. and {Ritchie}, B.~W. and {Smith}, K.~T.},
        title = "{On the nature of candidate luminous blue variables in M 33}",
      journal = {\aap},
         year = 2012,
        month = may,
       volume = {541},
          eid = {A146},
        pages = {A146},
          doi = {10.1051/0004-6361/201118440},
archivePrefix = {arXiv},
       eprint = {1202.4409},
 primaryClass = {astro-ph.SR},
       adsurl = {https://ui.adsabs.harvard.edu/abs/2012A&A...541A.146C}
}

@ARTICLE{Valeev2009,
       author = {{Valeev}, A.~F. and {Sholukhova}, O. and {Fabrika}, S.},
        title = "{A new luminous variable in M33}",
      journal = {\mnras},
         year = 2009,
        month = jun,
       volume = {396},
       number = {1},
        pages = {L21-L25},
          doi = {10.1111/j.1745-3933.2009.00654.x},
archivePrefix = {arXiv},
       eprint = {0903.5222},
 primaryClass = {astro-ph.CO},
       adsurl = {https://ui.adsabs.harvard.edu/abs/2009MNRAS.396L..21V}
}

@ARTICLE{Humphreys1980,
       author = {{Humphreys}, R.~M. and {Sandage}, A.},
        title = "{On the stellar content and structure of the spiral galaxy M33.}",
      journal = {\apjs},
         year = 1980,
        month = nov,
       volume = {44},
        pages = {319-381},
          doi = {10.1086/190696},
       adsurl = {https://ui.adsabs.harvard.edu/abs/1980ApJS...44..319H}
}

@ARTICLE{Maryeva2019,
       author = {{Maryeva}, Olga and {Viotti}, Roberto F. and {Koenigsberger}, Gloria and {Calabresi}, Massimo and {Rossi}, Corinne and {Gualandi}, Roberto},
        title = "{The History Goes On: Century Long Study of Romano's Star}",
      journal = {Galaxies},
         year = 2019,
        month = sep,
       volume = {7},
       number = {3},
          eid = {79},
        pages = {79},
          doi = {10.3390/galaxies7030079},
archivePrefix = {arXiv},
       eprint = {1909.08765},
 primaryClass = {astro-ph.SR},
       adsurl = {https://ui.adsabs.harvard.edu/abs/2019Galax...7...79M}
}

@ARTICLE{Chene2011,
       author = {{Chen{\'e}}, A. -N. and {Foellmi}, C. and {Marchenko}, S.~V. and {St-Louis}, N. and {Moffat}, A.~F.~J. and {Ballereau}, D. and {Chauville}, J. and {Zorec}, J. and {Poteet}, C.~A.},
        title = "{A 10-h period revealed in optical spectra of the highly variable WN8 Wolf-Rayet star WR 123}",
      journal = {\aap},
         year = 2011,
        month = jun,
       volume = {530},
          eid = {A151},
        pages = {A151},
          doi = {10.1051/0004-6361/201116567},
archivePrefix = {arXiv},
       eprint = {1104.5182},
 primaryClass = {astro-ph.SR},
       adsurl = {https://ui.adsabs.harvard.edu/abs/2011A&A...530A.151C}
}

@ARTICLE{Moffat1989,
       author = {{Moffat}, Anthony F.~J.},
        title = "{Wolf-Rayet Stars in the Magellanic Clouds. VII. Spectroscopic Binary Search among the WNL Stars and the WN6/7--WN8/9 Dichotomy}",
      journal = {\apj},
         year = 1989,
        month = dec,
       volume = {347},
        pages = {373},
          doi = {10.1086/168126},
       adsurl = {https://ui.adsabs.harvard.edu/abs/1989ApJ...347..373M}
}

@INPROCEEDINGS{Foellmi2002,
       author = {{Foellmi}, C. and {Moffat}, A.~F.~J.},
        title = "{Are Peculiar Wolf-Rayet Stars of Type WN8 Thorne-Zytkow Objects?}",
    booktitle = {Stellar Collisions, Mergers and their Consequences},
         year = 2002,
       editor = {{Shara}, Michael M.},
       series = {Astronomical Society of the Pacific Conference Series},
       volume = {263},
        month = jan,
        pages = {123},
          doi = {10.48550/arXiv.astro-ph/0607217},
archivePrefix = {arXiv},
       eprint = {astro-ph/0607217},
 primaryClass = {astro-ph},
       adsurl = {https://ui.adsabs.harvard.edu/abs/2002ASPC..263..123F}
}

@ARTICLE{Lazzarini2023,
       author = {{Lazzarini}, Margaret and {Hinton}, Kyros and {Shariat}, Cheyanne and {Williams}, Benjamin F. and {Garofali}, Kristen and {Dalcanton}, Julianne J. and {Durbin}, Meredith and {Antoniou}, Vallia and {Binder}, Breanna and {Eracleous}, Michael and {Vulic}, Neven and {Yang}, Jun and {Wik}, Daniel and {Gasca}, Aria and {Kuauhtzin}, Quetzalcoatl},
        title = "{Multiwavelength Characterization of the High-mass X-Ray Binary Population of M33}",
      journal = {\apj},
         year = 2023,
        month = aug,
       volume = {952},
       number = {2},
          eid = {114},
        pages = {114},
          doi = {10.3847/1538-4357/acdbc8},
archivePrefix = {arXiv},
       eprint = {2305.16390},
 primaryClass = {astro-ph.HE},
       adsurl = {https://ui.adsabs.harvard.edu/abs/2023ApJ...952..114L}
}

@ARTICLE{Kostenkov2020,
       author = {{Kostenkov}, A. and {Vinokurov}, A. and {Solovyeva}, Y. and {Atapin}, K. and {Fabrika}, S.},
        title = "{Modeling of Extended Atmospheres with Temperatures below 40000 K}",
      journal = {Astrophysical Bulletin},
         year = 2020,
        month = apr,
       volume = {75},
       number = {2},
        pages = {182-190},
          doi = {10.1134/S1990341320020078},
archivePrefix = {arXiv},
       eprint = {2007.02406},
 primaryClass = {astro-ph.HE},
       adsurl = {https://ui.adsabs.harvard.edu/abs/2020AstBu..75..182K}
}

@INPROCEEDINGS{Solovyeva2022,
       author = {{Solovyeva}, Y.~N. and {Kostenkov}, A. and {Dedov}, E. and {Vinokurov}, A.},
        title = "{Wind parameters of the new LBV in NGC1156}",
    booktitle = {The Multifaceted Universe: Theory and Observations - 2000},
         year = 2022,
        month = dec,
          eid = {49},
        pages = {49},
          doi = {10.48550/arXiv.2209.06012},
archivePrefix = {arXiv},
       eprint = {2209.06012},
 primaryClass = {astro-ph.SR},
       adsurl = {https://ui.adsabs.harvard.edu/abs/2022muto.confE..49S}
}

@ARTICLE{Najarro1997,
       author = {{Najarro}, F. and {Hillier}, D.~J. and {Stahl}, O.},
        title = "{A spectroscopic investigation of P Cygni. I. H and HeI lines.}",
      journal = {\aap},
         year = 1997,
        month = oct,
       volume = {326},
        pages = {1117-1134},
       adsurl = {https://ui.adsabs.harvard.edu/abs/1997A&A...326.1117N}
}

@ARTICLE{Maeder2000,
       author = {{Maeder}, A. and {Meynet}, G.},
        title = "{Stellar evolution with rotation. VI. The Eddington and Omega -limits, the rotational mass loss for OB and LBV stars}",
      journal = {\aap},
         year = 2000,
        month = sep,
       volume = {361},
        pages = {159-166},
          doi = {10.48550/arXiv.astro-ph/0006405},
archivePrefix = {arXiv},
       eprint = {astro-ph/0006405},
 primaryClass = {astro-ph},
       adsurl = {https://ui.adsabs.harvard.edu/abs/2000A&A...361..159M}
}

@ARTICLE{Dwarkadas2002,
       author = {{Dwarkadas}, Vikram V. and {Owocki}, Stanley P.},
        title = "{Radiatively Driven Winds and the Shaping of Bipolar Luminous Blue Variable Nebulae}",
      journal = {\apj},
         year = 2002,
        month = dec,
       volume = {581},
       number = {2},
        pages = {1337-1343},
          doi = {10.1086/344257},
       adsurl = {https://ui.adsabs.harvard.edu/abs/2002ApJ...581.1337D}
}

@ARTICLE{Binder2023,
       author = {{Binder}, Breanna A. and {Anderson}, Ashley K. and {Garofali}, Kristen and {Lazzarini}, Margaret and {Williams}, Benjamin F.},
        title = "{The spatial correlation of high-mass X-ray binaries and young star clusters in nearby star-forming galaxies}",
      journal = {\mnras},
         year = 2023,
        month = jul,
       volume = {522},
       number = {4},
        pages = {5669-5679},
          doi = {10.1093/mnras/stad1368},
archivePrefix = {arXiv},
       eprint = {2305.01802},
 primaryClass = {astro-ph.HE},
       adsurl = {https://ui.adsabs.harvard.edu/abs/2023MNRAS.522.5669B}
}

@ARTICLE{Poutanen2013,
       author = {{Poutanen}, Juri and {Fabrika}, Sergei and {Valeev}, Azamat F. and {Sholukhova}, Olga and {Greiner}, Jochen},
        title = "{On the association of the ultraluminous X-ray sources in the Antennae galaxies with young stellar clusters}",
      journal = {\mnras},
         year = 2013,
        month = jun,
       volume = {432},
       number = {1},
        pages = {506-519},
          doi = {10.1093/mnras/stt487},
archivePrefix = {arXiv},
       eprint = {1210.1210},
 primaryClass = {astro-ph.HE},
       adsurl = {https://ui.adsabs.harvard.edu/abs/2013MNRAS.432..506P}
}

@ARTICLE{Gvaramadze2008,
       author = {{Gvaramadze}, V.~V. and {Bomans}, D.~J.},
        title = "{Search for OB stars running away from young star clusters. I. NGC 6611}",
      journal = {\aap},
         year = 2008,
        month = nov,
       volume = {490},
       number = {3},
        pages = {1071-1077},
          doi = {10.1051/0004-6361:200810411},
archivePrefix = {arXiv},
       eprint = {0809.0650},
 primaryClass = {astro-ph},
       adsurl = {https://ui.adsabs.harvard.edu/abs/2008A&A...490.1071G}
}

@ARTICLE{Guo2024,
       author = {{Guo}, Yanjun and {Wang}, Luqian and {Liu}, Chao and {Wu}, You and {Han}, ZhanWen and {Chen}, XueFei},
        title = "{A Catalog of Early-type Runaway Stars from LAMOST DR8}",
      journal = {\apjs},
         year = 2024,
        month = jun,
       volume = {272},
       number = {2},
          eid = {45},
        pages = {45},
          doi = {10.3847/1538-4365/ad46f8},
archivePrefix = {arXiv},
       eprint = {2405.04750},
 primaryClass = {astro-ph.SR},
       adsurl = {https://ui.adsabs.harvard.edu/abs/2024ApJS..272...45G}
}

@ARTICLE{Massey2006,
       author = {{Massey}, Philip and {Olsen}, K.~A.~G. and {Hodge}, Paul W. and {Strong}, Shay B. and {Jacoby}, George H. and {Schlingman}, Wayne and {Smith}, R.~C.},
        title = "{A Survey of Local Group Galaxies Currently Forming Stars. I. UBVRI Photometry of Stars in M31 and M33}",
      journal = {\aj},
         year = 2006,
        month = may,
       volume = {131},
       number = {5},
        pages = {2478-2496},
          doi = {10.1086/503256},
archivePrefix = {arXiv},
       eprint = {astro-ph/0602128},
 primaryClass = {astro-ph},
       adsurl = {https://ui.adsabs.harvard.edu/abs/2006AJ....131.2478M}
}

@ARTICLE{Massey2007Halpha,
       author = {{Massey}, Philip and {McNeill}, Reagin T. and {Olsen}, K.~A.~G. and {Hodge}, Paul W. and {Blaha}, Cynthia and {Jacoby}, George H. and {Smith}, R.~C. and {Strong}, Shay B.},
        title = "{A Survey of Local Group Galaxies Currently Forming Stars. III. A Search for Luminous Blue Variables and Other H{\ensuremath{\alpha}} Emission-Line Stars}",
      journal = {\aj},
         year = 2007,
        month = dec,
       volume = {134},
       number = {6},
        pages = {2474-2503},
          doi = {10.1086/523658},
archivePrefix = {arXiv},
       eprint = {0709.1267},
 primaryClass = {astro-ph},
       adsurl = {https://ui.adsabs.harvard.edu/abs/2007AJ....134.2474M}
}

@ARTICLE{Bertin1996,
       author = {{Bertin}, E. and {Arnouts}, S.},
        title = "{SExtractor: Software for source extraction.}",
      journal = {\aaps},
         year = 1996,
        month = jun,
       volume = {117},
        pages = {393-404},
          doi = {10.1051/aas:1996164},
       adsurl = {https://ui.adsabs.harvard.edu/abs/1996A&AS..117..393B}
}

@ARTICLE{Elmegreen1997,
       author = {{Elmegreen}, Bruce G. and {Efremov}, Yuri N.},
        title = "{A Universal Formation Mechanism for Open and Globular Clusters in Turbulent Gas}",
      journal = {\apj},
         year = 1997,
        month = may,
       volume = {480},
       number = {1},
        pages = {235-245},
          doi = {10.1086/303966},
       adsurl = {https://ui.adsabs.harvard.edu/abs/1997ApJ...480..235E}
}

@ARTICLE{Wright2020,
       author = {{Wright}, Nicholas J.},
        title = "{OB Associations and their origins}",
      journal = {\nar},
         year = 2020,
        month = nov,
       volume = {90},
          eid = {101549},
        pages = {101549},
          doi = {10.1016/j.newar.2020.101549},
archivePrefix = {arXiv},
       eprint = {2011.09483},
 primaryClass = {astro-ph.SR},
       adsurl = {https://ui.adsabs.harvard.edu/abs/2020NewAR..9001549W}
}

@ARTICLE{Mapelli2011,
       author = {{Mapelli}, M. and {Ripamonti}, E. and {Zampieri}, L. and {Colpi}, M.},
        title = "{Dynamics of massive stellar black holes in young star clusters and the displacement of ultra-luminous X-ray sources}",
      journal = {\mnras},
         year = 2011,
        month = sep,
       volume = {416},
       number = {3},
        pages = {1756-1763},
          doi = {10.1111/j.1365-2966.2011.18991.x},
archivePrefix = {arXiv},
       eprint = {1105.0681},
 primaryClass = {astro-ph.CO},
       adsurl = {https://ui.adsabs.harvard.edu/abs/2011MNRAS.416.1756M}
}

@ARTICLE{Moeckel2010,
       author = {{Moeckel}, Nickolas and {Bate}, Matthew R.},
        title = "{On the evolution of a star cluster and its multiple stellar systems following gas dispersal}",
      journal = {\mnras},
         year = 2010,
        month = may,
       volume = {404},
       number = {2},
        pages = {721-737},
          doi = {10.1111/j.1365-2966.2010.16347.x},
archivePrefix = {arXiv},
       eprint = {1001.3417},
 primaryClass = {astro-ph.SR},
       adsurl = {https://ui.adsabs.harvard.edu/abs/2010MNRAS.404..721M}
}

@article{Efron1979bootstrap,
 author = {B. Efron},
 journal = {The Annals of Statistics},
 number = {1},
 pages = {1--26},
 publisher = {Institute of Mathematical Statistics},
 title = {Bootstrap Methods: Another Look at the Jackknife},
 urldate = {2025-08-29},
 volume = {7},
 year = {1979}
}

\end{document}